\documentclass[sigconf]{acmart}

\usepackage{multirow}
\usepackage{dirtytalk}
\usepackage{colortbl}
\usepackage{xcolor}
\usepackage{multirow}
\usepackage{multicol}
\usepackage{subfig}
\usepackage{soul}

\definecolor{new-green}{rgb}{0.104,0.667,0.229}

\newcommand{\rev}[1]{{\color{black}{#1}}}
\newcommand{\revn}[1]{{\color{black}{#1}}}
\AtBeginDocument{%
  \providecommand\BibTeX{{%
    \normalfont B\kern-0.5em{\scshape i\kern-0.25em b}\kern-0.8em\TeX}}}

\copyrightyear{2024} 
\acmYear{2024} 
\setcopyright{rightsretained} 
\acmConference[CHI '24]{Proceedings of the CHI Conference on Human Factors in Computing Systems}{May 11--16, 2024}{Honolulu, HI, USA}
\acmBooktitle{Proceedings of the CHI Conference on Human Factors in Computing Systems (CHI '24), May 11--16, 2024, Honolulu, HI, USA}
\acmDOI{10.1145/3613904.3642007}
\acmISBN{979-8-4007-0330-0/24/05}

\begin{document}

\title[Towards Designing a Chatbot]{Towards Designing a Question-Answering Chatbot for Online News: Understanding Questions and Perspectives}


\author{Md Naimul Hoque}
\email{nhoque@umd.edu}
\affiliation{%
  \institution{University of Maryland, College Park}
  \state{MD}
  \country{USA}
}

\author{Ayman Mahfuz}
\email{aymanmahfuz27@utexas.edu}
\affiliation{%
  \institution{University of Texas at Austin}
  \state{TX}
  \country{USA}
}

\author{Mayukha Kindi}
\email{mkindi@umd.edu}
\affiliation{%
  \institution{University of Maryland, College Park}
  \state{MD}
  \country{USA}
}

\author{Naeemul Hassan}
\email{nhassan@umd.edu}
\affiliation{%
  \institution{University of Maryland, College Park}
  \state{MD}
  \country{USA}
}

\renewcommand{\shortauthors}{Trovato and Tobin, et al.}

\begin{abstract}
\rev{ 
   Large Language Models (LLMs) have created opportunities for designing chatbots that can support complex \textit{question-answering} (QA) scenarios and improve news audience engagement. However, we still lack an understanding of what roles journalists and readers deem fit for such a chatbot in newsrooms. 
   To address this gap, we first interviewed six journalists to understand how they answer questions from readers currently and how they want to use a QA chatbot for this purpose. To understand how readers want to interact with a QA chatbot, we then conducted an online experiment (N=124) where we asked each participant to read three news articles and ask questions to either the author(s) of the articles or a chatbot. By combining results from the studies, we present alignments and discrepancies between how journalists and readers want to use QA chatbots and propose a framework for designing effective QA chatbots in newsrooms.}
   %

   % Findings from the two studies suggest there is a high-level alignment between how journalists and readers want to use a QA chatbot. Both parties see the chatbot as a \textit{mediator} between them and predominantly want the chatbot to answer questions seeking \textit{facts} and \textit{details} while wanting to engage between themselves for questions seeking  \textit{subjective interpretation}. We conclude by proposing a framework for designing a QA chatbot in online news.}
    %
   % Without this knowledge, we may not be able to design chatbots that align with readers' and newsrooms' expectations. 
   %
   % The interviews suggest journalists want to use the chatbot for answering factual and redundant questions from the readers, while retaining authority to answer questions expecting explanations and opinions.
    % \del{We removed this sentence.}
    % \rev{We revised this text.}
    % \add{We added this text.}
    % \marginpar{\footnotesize We could also add margin notes, like this one, to explain changes.}
\end{abstract}





\maketitle
\section{Introduction}

Journalists and reporters carefully integrate information and interpretation in a news article to inform readers about a topic, event, or phenomenon in a~\textit{balanced} and~\textit{objective} manner~\cite{mencher1997news, campbell1997journalistic, maras2013objectivity, starkey2017balance}. Nonetheless, prior research shows that oftentimes readers are left with questions for journalists/authors after reading news~\cite{castellano2020behind, weber2014discussions}. Readers use various means such as news comment sections, email, and social media direct messages to ask their questions~\cite{DBLP:conf/cscw/DiakopoulosN11, martin2021negotiating}. However, due to many reasons such as the abundance of questions, lack of time, and absence of direct incentives, journalists cannot respond to the queries~\cite{cohen2019work, monteiro2016job}. On the other hand, both journalists and newsrooms have an interest in engaging with the audience/readers, to learn what aspects of their news raised curiosity and questions in readers' minds~\cite{meier2018audience, nelson2021next}, and to understand if readers are questioning newsroom's journalistic integrity~\cite{masullo2017building}. This knowledge is critical for learning what attracts readers to news content, targeting the right audience for news recommendations, building trust with the audience and so on~\cite{li2010user, wu2020user}. According to Meier et al.~\cite{meier2018audience},- ``a paradigm shift away from a `\textit{lecturing}' approach to a `\textit{dialogue}' approach is a key factor for journalism in a post-truth era''. \revn{Thus}, there is a clear interest from both readers' and newsrooms' perspectives to engage with each other in a~\textit{dialogue} fashion through question-answering~\cite{meier2018audience, pena2021social}. 
% A chatbot that can automatically answer questions from readers can potentially be useful in such scenarios~\cite{10.1145/3543829.3544530, DBLP:conf/iui/XiaoLZG023}.

With the rise of Large Language Models (LLMs)~\cite{brown2020language}, like any other industry, news media industry is also investigating how Artificial Intelligence (AI) and LLMs can be integrated with the news production, distribution, audience engagement, and other related processes to improve efficiency, increase trust and accountability, and open new possibilities~\cite{veglis2019embedding, hong2020utilizing, veglis2019chatbots}. For instance, BBC has started experimenting with how we can leverage bot technologies to reach new audiences on messaging platforms and social media~\cite{bbcnewslabs, bbc_scripting}. They have developed a prototype of an in-article chatbot to help less engaged audiences understand big news stories~\cite{bbcnewslabsinarticle}. Other media outlets have started experimenting with Meta (previously known as Facebook) chatbots~\cite{zhang2021informing}. While these experiments are ongoing individually in different newsrooms, \rev{facilitating question-answering (QA), one of the primary modes of interaction and dialogue in chatbots~\cite{10.1145/3173574.3173577}, remains a challenge for newsrooms~\cite{trischler-etal-2017-newsqa, verge}. While researchers are working towards mitigating technological limitations such as the lack of truthfulness and existence of biased and formulaic text in LLM responses~\cite{DBLP:journals/corr/abs-2112-04359, Epstein_2023}, we argue that it is also important to take a systematic approach to understand how journalists and readers want to use a chatbot for question-answering and what roles they deem fit for such a chatbot. For instance, \textit{are journalists comfortable with a chatbot answering readers' questions about their written news article? What types of questions \revn{do readers} expect the chatbot to answer? And what types of questions \revn{do readers} expect the authors to answer?}
Without answering these questions, we may end up designing QA chatbots that are detached from readers' and newsrooms' needs. }

% We believe our research will inform the design of future LLM-powered chatbots.

\rev{This paper takes a \textit{formative} step towards designing LLM-powered chatbots that will be able to answer open-ended questions from readers, \revn{reduce journalists' burden} to answer a large number of queries, and at the same time elevate communication between readers and journalists. We call such chatbots as QA chatbots. For the rest of the paper, we use the term QA chatbots and chatbots interchangeably to refer to the chatbots that are able to answer questions from news readers, unless specified otherwise. While we only focus on question-answering in the context of LLMs, in a real-world setting, like many LLM-powered chatbots (e.g., ChatGPT), a QA chatbot (or simply a chatbot) will likely have other forms of interactions (e.g., recommending news). } To achieve our research agenda, we conducted two studies in this paper. First, we interviewed six professional journalists to understand how journalists currently answer questions from readers and, more broadly, interact with readers. The interviews revealed that most journalists now obtain feedback and answer questions from readers through one-to-one conversations over email, direct messages, or in-person meetups, except for a few member gatherings organized by the news outlets. All participants were enthusiastic about \rev{using a QA chatbot as a \textit{mediator} between themselves and readers. They want the chatbot to answer factual and redundant questions from readers while want the chatbot to direct questions seeking subjective interpretation towards them.}   

 \rev{We subsequently conducted a between-subject experiment on Amazon Mechanical Turk (MTurk) with 124 participants with two goals: (1) understanding to what extent readers' perspectives and questioning patterns match with how journalists see the functionalities of a QA chatbot; and (2) understanding how a QA chatbot might modulate readers' questioning patterns. The second goal is motivated \revn{by} prior research that shows that chatbots can significantly modulate user behaviour~\cite{10.1145/2858036.2858288, doi:10.1080/10447318.2020.1841438, 10.1145/2901790.2901842}.} In the study, participants were instructed to read three articles from three different domains (health, politics, and environment) and ask questions to the authors of the articles or a chatbot. We analyzed their responses using a grounded theory approach~\cite{glaser2017discovery} and a set of quantitative measures. Our findings suggest that readers mostly asked questions on two broad categories: \textit{Information}, questions that seek short-form factual and long-form details; and \textit{Interpretation}, questions that seek explanation and opinion. We found that readers' questioning patterns significantly differ depending on the receiver - authors vs. chatbot.  Participants asked questions seeking \textit{facts} more frequently to the \textit{chatbot} than to the \textit{authors}.  In contrast, they asked questions seeking \textit{explanation} and \textit{opinion} with multiple facets more frequently to \textit{authors} than to the \textit{chatbot}.
 
 \rev{These results indicate that readers' perspectives about the role of QA chatbots match with journalists' \revn{views} to some extent. Readers too predominantly want the chatbot to answer factual questions while seeking to engage in subjective conversations with journalists. }
 \rev{Despite this alignment, we found evidence that a significant number of readers do not share the view that chatbots should only answer factual questions. For instance, while not as frequently as to the authors, readers did ask a considerable amount of questions seeking subjective interpretation to the chatbot.
 %
 % In fact, 45 out of 62 participants who asked questions to the chatbot, seek answers to both factual and interpretive questions, albeit factual questions being the frequent category. 
 %
 % Similarly, readers do not want authors to completely abandon answering factual questions. A considerable amount of readers still seek answers to factual questions from the authors.
 Similarly, a considerable number of readers wanted authors to answer factual as well as subjective questions.
 Finally, we found that chatbots can modulate readers' behavior significantly which researchers and news organizations should be aware of. For example, articles having lower \textit{perceived quality} motivated readers to be critical and question journalistic integrity when the authors were present in the loop. However, when the chatbot was present, readers did not engage on the same level, even when the perceived quality of the news article was low.}

Overall, these results indicate that researchers and news organizations need to understand journalists' and readers' perspectives and expectations thoroughly \rev{before determining the functionalities of a QA chatbot}. The results also highlight the need to devise policies around chatbots and communicate the policies to journalists and readers effectively. We conclude this paper by proposing a framework for devising a policy for designing QA chatbots for online news. In summary, our contributions are as follows: (1) An interview study with six journalists to understand their current practices and challenges for answering questions from readers and how they perceive the role of an intelligent chatbot in this regard; (2) An online experiment with \rev{124 participants} to understand what questions readers typically ask about a news article and how the questions change if a chatbot is present at the receiving end, instead of the authors; and (3) A design framework for developing a chatbot policy, informed by the interviews, online experiment, and literature review.

\section{Background and Related Work}
% \textcolor{red}{Still underdeveloped. Comments welcome.}

% This work advances our knowledge of readers' 
In this section, we discuss prior works on understanding communication between journalists and readers. We also review Question-Answering (QA) based chatbot technology.

\subsection{Audience Engagement and Chatbots}
Audience or reader engagement plays an important role in news production, dissemination, and consumption~\cite{doi:10.1080/17512786.2020.1722729}. Readers often want to connect to authors for providing compliments and encouragement, engaging in intellectual discussion, asking follow-up questions, or for pointing out shortcomings~\cite{TANDOC2017149, DBLP:conf/cikm/RischK20, doi:10.1177/1464884915579332}. Journalists want to connect with readers for occupational duties, community engagements, obtaining new directions, or self-improvement~\cite{DBLP:conf/chi/AtrejaSJP23, DBLP:conf/chi/WangD21, williams2010limits, von2018sourcing}. As most news sites became online in early 2000, communication between journalists and readers has evolved from one-to-one interaction to various forms of online activities.  For example, the comment section in news sites had shown promises for readers to provide feedback to the journalists and incite community discussion~\cite{DBLP:conf/cscw/DiakopoulosN11, DBLP:conf/icwsm/RischK20}. However, several concerning issues such as the use of abusive language, spam, and a lack of manpower for moderation have contributed to comment sections becoming virtually non-existent in major news sites~\cite{journalmedia2040034}. 

Social media has drastically changed audience engagement and news consumption paradigm (e.g., Twitter)~\cite{doi:10.1080/17512786.2020.1722729, doi:10.1080/17512786.2019.1585198}.
About half of the U.S. population now access news from social media~\cite{paw}. Thus, news media are no longer the authoritative sources of news~\cite{doi:10.1177/1461444817750396}. With this shift in dynamics, news organizations are exploring different ways to adapt to social media environments and engage audience~\cite{su12166515}. One of the promising ways is designing chatbots in social media~\cite{su12166515, doi:10.1080/21670811.2015.1081822}. For example, several large international news organizations such as ABC News, NBC News, and BBC News have chatbots in Facebook messenger~\cite{zhang2021informing}. These chatbots, including others in this area~\cite{doi:10.1080/21670811.2015.1081822, fi12070109, zhang2021informing}, largely focus on supporting fact-checking, news dissemination, and recommendation. 
% Question-answering, on the other hand, has remained an interaction paradigm between readers and authors (e.g., journalists)~\cite{DBLP:conf/cikm/RischK20}.

However, similar to how technology has shaped audience engagement throughout history, it is likely that LLMs will influence the design of chatbots for news outlets~\cite{DBLP:conf/cui/NordbergG23}. LLMs can generate free-form text for a query, opening up opportunities for designing chatbots for complex QA tasks. At the same time, LLMs can hallucinate and provide wrong or even non-existent information to users~\cite{DBLP:journals/corr/abs-2112-04359}. Thus, it is imperative that we study what questions the journalists and audience want a chatbot to answer and align the design of the LLM-powered chatbot to match that expectation. We aim to address this gap.

\subsection{How End-users Perceive Chatbots?}
Chatbots are ubiquitous in many domains now, from  Facebook Messenger to most online services, customer-facing platforms~\cite{DBLP:journals/interactions/BrandtzaegF18, zhu-etal-2018-lingke}, educational platforms~\cite{10.1007/978-3-319-39583-8_27}, and health-related services~\cite{info:doi/10.2196/mental.7785, 10.1145/3589959}. 
Chatbot has become a popular choice for solving many user-centric problems in HCI~\cite{10.1145/3173574.3174178, 10.1145/3313831.3376785, 10.1145/3313831.3376131, 10.1145/3589959, DBLP:journals/interactions/BrandtzaegF18, 10.1145/3173574.3174042, 10.1145/3544548.3581109, 10.1145/3173574.3173632}.
Beyond developing chatbots, HCI research has a long history of studying how chatbots impact user behavior, interactions, and expectations. For example, Luger et al.~\cite{10.1145/2858036.2858288} found user expectations to be dramatically different than the capabilities of the chatbot. Liao et al.~\cite{10.1145/2901790.2901842} found that preference for humanized social features varies from user to user based on the underlying task requirements. Several other studies reported that user preference and satisfaction largely vary among individuals~\cite{10.1145/3173574.3173577, 10.1145/3196709.3196735}.

Given the prior evidence of chatbots mediating user expectations and behavior in other domains, we believe it is essential to study how news readers perceive the capabilities of a chatbot and how that impacts their expectations from the chatbot. To study that, we explore what questions readers would ask a chatbot in the context of question-answering, the primary form of interaction with chatbots~\cite{10.1145/3173574.3173577}.

\subsection{Question-Answering in NLP}
Question-answering (QA) is a core NLP task~\cite{rajpurkar-etal-2016-squad, rajpurkar-etal-2018-know}. Extractive QA or Reading Comprehension is the simplest form of QA task, where a model answers a question from a context~\cite{rajpurkar-etal-2018-know}. There are many challenges to the Reading Comprehension task. For example, a question can be \textit{Unanswerable} because of \textit{unverifiable presupposition}~\cite{kwiatkowski-etal-2019-natural}. QA systems that allow answers for \textit{unverifiable presupposition} may produce inappropriate answers~\cite{51002}. More complex QA tasks include generating free-form answers from a context or generating free-form answers from an open domain. Another way to categorize QA systems is to consider whether the system is rule-based, ML-based, or hybrid~\cite{10.1145/3313831.3376131}. Rule-based systems can provide exact answers to pre-defined questions but do not scale to a large volume of questions. ML-based systems have more language understanding but require a large amount of data and model training. Hybrid systems combine both rule-based answers and ML to answer questions~\cite{DBLP:conf/iui/XiaoLZG023}.

% LLMs such as ChatGPT largely operate in a conversational manner. Since these models have a large knowledge base, they can answer a wide variety of questions on diverse topics and tasks. However, LLMs can hallucinate and provide wrong or even non-existent information to users~\cite{DBLP:journals/corr/abs-2112-04359}. Additionally, LLMs are expensive to train, thus their knowledge base is difficult to keep up to date. Thus, researchers are advocating for hybrid chatbots that can utilize AI power while integrate expert opinions~\cite{DBLP:conf/iui/XiaoLZG023}. 

% Our work is built upon this premise. We believe 

 % found that user interactions vary whether the user is seeking information or looking for playful conversation. The interactions can be modeled to determine user satisfaction while adapting to different conversational styles. Jain et al.~\cite{} found that users prefer chatbots that provid either a ``human-like'' natural language conversation ability, or an engaging experience that exploited the familiar turn-based messaging interface.

% Our findings indicate that the 
% ~\cite{doi:10.1080/10447318.2020.1841438, DBLP:journals/interactions/BrandtzaegF18, 10.1145/3173574.3173577}

\begin{table*}[t]
    \centering
    \begin{tabular}{ccccp{4.5cm}cc}
    \toprule
    \rowcolor{gray!10} 
    \textbf{Id} & \textbf{Gender} & \textbf{Age} & \textbf{Education}  & \textbf{Experience} & \textbf{Yrs Exp} & \textbf{\rev{News Organization}} \\ 
    \midrule
    
    P1 & Female & 25-34 & Bachelor & Reporter (business, health, and politics) & 8 & \rev{National}  \\
    \rowcolor{gray!10}

    \rowcolor{gray!10}
    P2 & Female & 35-44 & Masters & Reporter (health) & 12 & \rev{Local} \\

    P3 & Male & 35-44 & Doctorate & Investigative journalist, \rev{data} scientist, and lecturer & 16 & \rev{Global}  \\
    
    \rowcolor{gray!10}
    P4 & Male & 45-54 & Bachelor &   News applications developer, database reporter, and lecturer & 25 & \rev{Global}  \\

    P5 & Female & 55-64 & Doctorate & Investigative journalist, science reporter, and professor & 40 & \rev{Global}  \\

    \rowcolor{gray!10}
    P6 & Female & 18-24 & Bachelor &  Reporter and  \rev{writer} & 4 & \rev{National}  \\
    
    \bottomrule
    \end{tabular}
    \caption{\textbf{Participant demographics for the formative study.}
    }
    \Description{A table with six rows and six columns, showing the demographic information of six participants. The columns are ID, Gender, Age, Profession, Experience, and Years of Experience.}
    \label{tab:participant_formative}
\end{table*}

\section{Study with Journalists}
\label{sec:formative}
We conducted IRB-approved semi-structured interviews with six journalists to understand how journalists interact with readers, answer questions, their challenges, and the opportunity for a QA chatbot in the process. \rev{The interviews took place in between January and February  2023}.

\subsection{Participants}

We contacted professional journalists who also work as faculty at our university's journalism school. We also reached out to journalists at local and national news outlets. Six journalists agreed for the interviews. Their information is given in Table~\ref{tab:participant_formative}. Participation was voluntary with no compensation.

\subsection{Method}

% Interviews were semi-structured and lasted around an hour. 

We conducted two interviews in person and others over Zoom.
One author of this paper conducted the interviews while another author took notes.
Each interview was divided into three parts.
First, after gathering informed consent, we asked participants to describe how they currently interact with readers, answer questions from readers, and manage conversations with readers. 
After that, we asked participants about any challenges they faced in the process. 
Finally, participants brainstormed with the authors about how a chatbot can enrich journalist-reader communication. Each session lasted around 1 hour. The semi-structured questionnaire is available in our \textcolor{cyan}{\href{https://osf.io/9nm7r/}{OSF repository}}.

\subsection{Data Collection and Analysis}
We recorded audio for all interviews and made anonymized transcripts for each.
Two authors of this paper analyzed the interview data, following a thematic analysis process~\cite{braun2006using}.
Throughout the analysis, we refined themes and relevant passages required to support the themes.
We present the findings in the next section.

\subsection{Results}

\subsubsection{How do journalists communicate with readers?}
Participants mentioned several ways to interact with readers. Most interactions happen asynchronously, over emails and social media (P1-P6). All participants valued communication with readers (P1-P6). However, they often do not have enough time to interact with the readers (P1, P4, P6). 

Participants mentioned comment sections are non-existent in their newsrooms, except for a few special reports (P3-P5). Newsrooms do not have enough workforce to moderate comment sections (P5). Some newsrooms periodically invite a subset of readers (often subscribers) to virtual or in-person gatherings for feedback and discussion (P1). When asked about chatbots, P2-4 mentioned that they are aware of several deployed chatbots from news organizations. However, the chatbots are not used for answering questions, but rather for recommending articles to readers. 

\subsubsection{Challenges for journalists}
Journalists mentioned that maintaining fruitful discussions with readers at scale is their main challenge (P1-P6). They are often rushed from one news to the next one. As a result, it is difficult for them to engage with readers on a previous article. Even when they interact with readers, they fear abuse and threats, especially if the article's topic is polarized (P1, P5).

\subsubsection{Opportunities and requirements for a chatbot}
We asked participants to comment on the opportunities for a chatbot in this domain. All our participants were aware of the recent bloom in LLMs. P3 and P4 mentioned that some newsrooms already have internal talks on how to use these technologies responsibly. P2 referred to BBC's recent efforts in designing chatbots~\cite{bbc_scripting}. 

However, participants mentioned several challenges for AI-based chatbots. First, participants think it is essential to decide the purpose of the chatbot and its extent of engagement with readers. For example,  P3 and P1 said, 
\begin{quote}
    \textit{``I wouldn't mind AI answering some factual questions from readers. I sometimes get redundant questions from readers. AI could be helpful there. However, I do not think I will use it if the reader is seeking a deep conversation or asking critical questions.'' (P3)}
    
    \textit{``A chatbot can be helpful to both audience and journalists. It can work as a learning medium for readers. However, I do not want to be completely separated from my audience. I would also use it cautiously. I wouldn't be comfortable knowing that an AI might misinterpret my writing and propagate that to the readers.''(P1)}
\end{quote}
 
  Finally, according to P4, it is a ``three-way street,'' and it is important to find the ``balance'' and ``mechanisms'' for journalists, readers, and AI to interact with each other. 

% they might miss out on important feedback and followup directions if they do not interact with readers at all (P1-P5). Second, P1 and P3 mentioned AI should not try to provide opinion to readers on sensitive topics, rather should be used to reduce burden from journalists by answering redundant questions from readers.  
\section{Study with Audience/Readers}

The interviews highlighted the challenges for journalists to maintain conversations with readers. They often do not have enough time and opportunities for fruitful conversations or answer questions from readers. \rev{All journalists were enthusiastic about using a chatbot as a mediator between them and readers. They wanted to stay connected with readers and answer important questions while being able to use chatbots for factual and redundant questions.}

\rev{While the perspective of journalists is clear from the interviews, to design an effective QA chatbot, we still need to know how readers, the other significant stakeholder, perceive chatbots. More importantly, since journalists want to use the chatbot as a mediator, we wanted to know to what extent readers' questioning patterns match with this view. We determined that conducting a study to understand how readers want to route questions between the authors and \revn{the} chatbot would be essential to achieve this. We also wanted to understand how a chatbot might influence the questioning patterns \revn{of} readers.  Prior research suggests chatbots can significantly modulate user behaviour~\cite{10.1145/2858036.2858288, doi:10.1080/10447318.2020.1841438, 10.1145/2901790.2901842}. Thus, we seek to answer the following two research questions: }

\begin{itemize}
    \item[\textbf{RQ1:}] What are the types of questions readers would ask about the news?
    \item[\textbf{RQ2:}] How do the questions differ when readers ask the questions to a chatbot in comparison to the actual authors?
\end{itemize}

% To facilitate that, we need a deeper understanding of the questions that readers ask frequently. On the other hand, investigating readers' perspectives on using a chatbot for answering questions is equally important. 

\rev{In addition, we wanted to study the effect of confounding factors, such as readers' perceived quality of the article~\cite{doi:10.1177/1464884916641269} and preference for news outlets with political alignment~\cite{10.1257/aer.20191777}, on the process. Thus, our second set of RQs are as follows:}

\begin{itemize}
    \item[\textbf{RQ3:}] How would the perceived quality of the articles influence the questions?
    \item[\textbf{RQ4:}] How would the readers' personal preferences for news outlets influence the questions?
\end{itemize}

To answer the RQs, we decided to conduct a crowdsourced experiment. \rev{While \revn{an interview study} with readers could also be useful here, we decided that a crowdsourced experiment is better suited here since it would enable us to understand the perspectives of a larger population, engage readers in the task of concern (asking questions), and help us answer the RQs.} This section describes the study design and protocol. All study materials, including news articles and the source code for the study interface, are available in our \textcolor{cyan}{\href{https://osf.io/9nm7r/}{OSF repository}}. \rev{The study took place in June 2023.}

\subsection{Study Conditions}
\rev{We conducted a between-subject study where we asked participants to read news articles and ask questions to the following two entities (conditions):}

\begin{itemize}
    \item[\textbf{A.}] \textbf{Authors:} Participants were prompted to ask questions to the authors of an article. \rev{Participants received the following prompt: \textit{``Consider you have the option to directly ask questions to the authors of this article about the article and get answers from them. Please list any questions you would ask the authors if you had the opportunity to communicate with them online. You can ask for clarifications, followup information, opinions, or any other questions that you think are relevant to the article.''}}

    \item[\textbf{C.}] \textbf{Chatbot:} Participants were prompted to ask questions to a chatbot. \rev{Participants received the following prompt: \textit{``Consider there is an automated chatbot that can answer your questions
    about this article. Please list any questions you would ask the chatbot if you had the opportunity to communicate with it online. You can ask for clarifications, followup information, opinions, or any other questions that you think are relevant to the article.''}}
\end{itemize}

\rev{While many chatbots can converse with humans (e.g., ChatGPT), the chatbot in our study (\textbf{C}) did not feature any response or interactions. There are three reasons for that. First, we see this study as a formative step to design future chatbots. We wanted to understand user needs and perceptions before developing the chatbot. Thus, while we acknowledge that an actual chatbot will likely influence readers' perceptions, we believe we would be able to study that influence once we design a chatbot informed by this study. Second, current LLMs have several limitations such as providing information that is wrong or even non-existent, biased text generation, and so on~\cite{Epstein_2023}. These problems can significantly impact participants' experience in the study which would be difficult for us to control in an open-ended setting such as ours.  Finally, it is unlikely that we would be able to recruit journalists in the study who would be able to manage time to answer questions on the fly for the Authors condition (\textbf{A}). Thus, an actual chatbot capable of replying will be biased against the Authors (\textbf{A}) condition.}

% Our goal in this project is to understand what kinds of questions people want to ask an automated chatbot in response to reading a news article. You are eligible to participate if you are 18 or more and can read news articles written in English.

 Half of the participants asked questions using \textbf{A} while the other half used \textbf{C}. Our goal was to collect ~\rev{375 questions} for each condition. This is motivated by prior work~\cite{DBLP:conf/cikm/RischK20}, which used 1500 comments in online news to derive a taxonomy of engaging comments. However, only 3 out of 12 classes in that taxonomy are for questions, equaling around 375 data on average for deriving the taxonomy for questions. 
 % However, contrary to our study, their study did not include any conditions. 
 \rev{We decided that around 750 questions in total (750/2 = 375 questions per condition) would enable us to capture the differences between \textbf{A} and \textbf{C}, if any.}

% We present the results in the next section.
\subsection{Participants}
We recruited \rev{126 participants} from Amazon Mechanical Turk (MTurk). \rev{We excluded 2 participants due to incomplete responses, resulting in 124 participants}. We decided on the number of participants by considering the estimated responses required for answering the RQs (around \rev{375 per condition}). We considered that crowd workers typically have a short attention span and perform well on short repetitive tasks~\cite{DBLP:conf/cscw/KitturNBGSZLH13}. Thus, we decided to ask participants to ask only two questions per article. Participants repeated this task for 3 separate news articles. \rev{We determined the number of articles to read, along with other study variables, by conducting a pilot study with 10 undergraduate and graduate students at our university.  Overall, participants found the process of reading 3 articles to be easy and quick. This also allowed us to assign one article from three different domains (Health, Politics, and Environment) to each participant (see \autoref{sec:articles}) and curate a wide range of questions. Finally, we anticipated that we would be able validate the responses from crowdworkers confidently if we could measure responses across three articles.}
Each session resulted in 3 x 2 = 6 questions. \rev{Thus, we recruited 375/6 = 62.5 or 63 participants per condition and 126 participants in total. We conducted Monte Carlo simulation to ensure that the Structural Equation Model (SEM) proposed in the result section (\autoref{sec:sem}) has a power of over 0.8 (0.82 to be exact). }

The details of the participants are provided in Figure~\ref{fig:participant} \revn{(Appendix A)}. MTurk has several embedded functionalities to control the quality of participants. We required participants to be located in the United States (US), have at least 50 approved tasks, and be \rev{Master workers, a qualification assigned to workers who have demonstrated a high rate of success in completing a wide range of tasks}. 
On average, it took \rev{29 minutes} for participants to complete the study. Considering the minimum wage standards in our US State, we compensated each participant \$8 for their time.

\begin{figure*}
    \centering
    \includegraphics[width=0.95\textwidth]{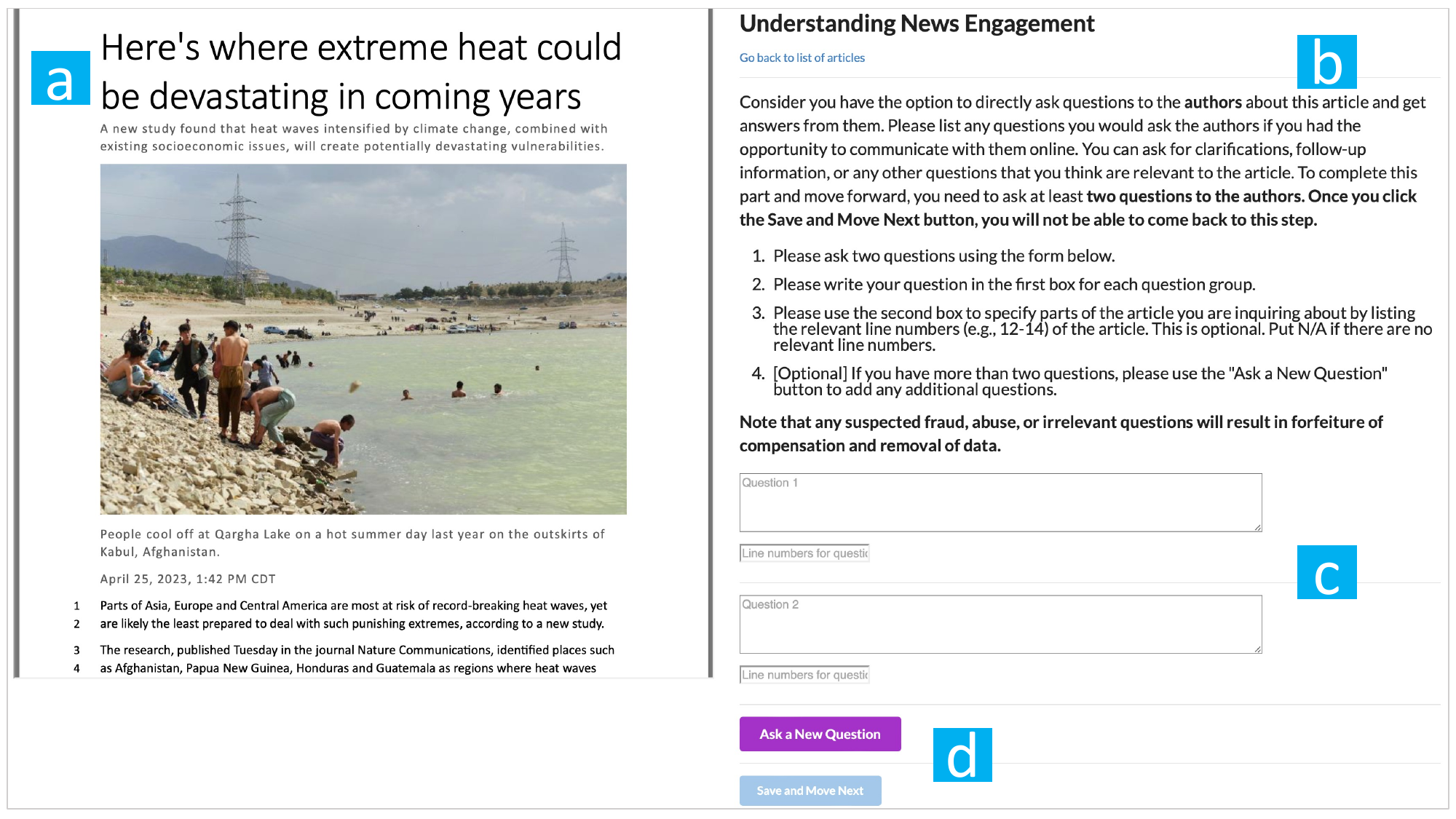}
    \caption{\textbf{Web interface for the study.} Participants used this interface to read three anonymous articles and ask questions to the authors or a chatbot. The screenshot shows the interface for an article. (a) A PDF reader for reading the article. (b) Instruction for completing the task. (c) Input boxes for writing the questions. (d) Buttons to write a new question (optional) and move forward (enabled only when a participant provides two questions). }
    \label{fig:interface}
    \Description{This figure shows a web interface, on the left it shows a new article, on the right it shows a paragraph of text content, some input boxes, and buttons.}
    
\end{figure*}

\subsection{News Articles and Collection Method}
\label{sec:articles}
% We will avoid long articles, instead choose articles shorter than 600 words, so that crowd workers can quickly read them.
We curated 15 news articles from three separate domains (five articles per domain): Health, Politics, and Environment. We chose Health as there has been an increased public interest in health-related reports due to COVID-19. There are reports of increased misinformation and pseudoscience in this domain. Similarly, politics and environmental issues such as climate change are contentious areas. People tend to have divergent opinions about political and environmental issues. We anticipated these domains would result in a wide range of questions, helping us answer the RQs. 

% We curate the articles from a wide variety of sources (e.g., New York Times, Fox News) to introduce diversity and reduce chances of sampling bias.

To curate the articles, we identified the news outlets we intended to source articles from. Our objective was to select outlets that spanned the political spectrum, ensuring that the articles encompassed diverse biases. To achieve this, we consulted the AllSides media bias website~\cite{allsides} to identify organizations that could be categorized into five distinct groups: far left, leaning left, center, leaning right, and far right.
Subsequently, we performed a search for articles followed by one of the aforementioned categories (e.g., "Environment articles, far left-leaning"). The first page of search results yielded various articles from multiple outlets. We then examined the first article and cross-referenced the outlet it originated from with AllSides to confirm that the outlet aligned with the intended category. We selected that article if the outlet's classification was consistent with the desired category. If not, we proceeded to evaluate the next article in the search results. All articles chosen for this study were sourced from the first page of the Google search results. We also intentionally avoided long articles, so that crowd workers could quickly read them. The average length of the 15 selected articles was around 650 words.

By employing this systematic approach to article selection, we ensured that our study incorporated a wide variety of perspectives and minimized potential biases in the articles analyzed.
The final counts for the source news outlets are--
Reuters: 1, BBC News: 2, NBC: 1, NPR: 1, AP News: 1,
CNN: 1, NY Times: 2, Washington Times: 1, NY Post: 2, Fox News: 2, Daily Mail: 1. To avoid potential bias towards a news organization, we removed all identifiable information such as Author's name from the articles and replaced reference to the news organizations name with generic names. \rev{For each participant in the study, we randomly chose one article from each domain. Figure~\ref{fig:article_dist} \revn{(Appendix A)} shows the number of times the articles were read by the participants. }

\subsection{Study Interface}
We developed a web interface for conducting the study (Figure~\ref{fig:interface}). Participants used the interface to provide consent, read instructions and three news articles, and provide two questions for each article. We used \textsc{Python} in the backend for web services with \textsc{JavaScript} in the front end for supporting user interaction. We used \textsc{MongoDB} to save responses from participants. 
% We include several other screenshots of our interface as supplemental materials.
 \begin{table*}[t]
    \centering
    \begin{tabular}{c|m{1.7cm}m{5cm}m{5cm}}
    \toprule
    \rowcolor{gray!10} 
    & \textbf{Category}  & \textbf{Description} & \textbf{Example} \\ 
    \hline

    % \rowcolor{gray!10}
    \multirow[b]{3}{1.5cm}{Information (\rev{619})} & Fact (\rev{211}) & Expects a short, objective answer to \textit{When?} or \textit{Where?} & When was the site first used as a clothing dump? \\
    \cline{2-4}

    & Details (\rev{302}) & Expects details or more information about a topic that could not be answered with short factual information & What are the benefits of combustion engines that run on ammonia? \\
    \cline{2-4}

    % \rowcolor{gray!10}
    & Evidence (\rev{56}) & Expects evidence for a claim in the article  & Is there any evidence as to the impartiality of the chief of the USDA? \\
    \cline{2-4}
    
    & Closed-ended (\rev{50}) & Expects a simple \textit{yes} or \textit{no} answer & \rev{Was this announcement televised?} \\
    \hline
    
    \multirow[b]{2}{1.7cm}{Interpretation (\rev{266})} & Opinion (\rev{69}) & Expects a long, subjective answer to the question \textit{What do you think about...?} \textit{What would you suggest?} & What do you think the effect of having multiple COVID-19 vaccine boosters will have on people's perception of the COVID-19 vaccine and vaccines in general?  \\
    \cline{2-4}

    % \rowcolor{gray!10}
    & Explanation (\rev{197}) & Expects an answer to the questions \textit{Why?} \textit{How can I...?} in the form of a (long) explanation & Why do the holidays throw outbreak data off by so much? \\
    \hline

    Others (\rev{9}) & --- & Questions that do not match any of the above categories & Can you summarize the movement of the migrants, according to the article? \\

    \bottomrule
    \end{tabular}
    \caption{\textbf{A taxonomy of questions.} We identified 3 high-level categories (Information, Interpretation, Others) and 7 sub-categories/types from our analysis of the questions. The numbers represent counts for each category.
    }
    \label{tab:types}
    \Description{A table with three rows and four columns. The first row has four sub-rows for the second, third, and fourth columns. The second row has two sub-rows for the second, third, and fourth columns. The third row has no sub-rows.}
\end{table*}

\subsection{Protocol}
We used the ``external HIT'' functionality on AMT, where the interface hosted on our server was accessible to the workers. Participants followed the following sequential workflow in the study:

% Therefore, our interface did not require the crowd workers to log in and provide personal information beforehand. 
 
 \begin{enumerate}
 \item \textbf{Read instruction and provide consent:}
 Participants read an overview of the study, including what they will be doing in the study and the requirements for completing the study. After reading the instructions, participants provided consent for the study. We notified participants that all responses will be validated by the research team and any suspected fraud or abuse will result in forfeiture of compensation and removal of data.

 \item \textbf{Read an article and ask questions (x3 articles):} \rev{ Participants completed the following four tasks for each article in the study. } 

 \begin{enumerate}
     \item \textit{Read the article:} Participants read the article in a PDF reader.
     \item \textit{Pass the reading comprehension test:} As a quality control step, we asked participants to answer two multiple-choice questions relevant to the article. Participants had access to the PDF in this step too. The questions were designed to test participants' comprehension of the article \rev{and to ensure that they do not move to the next step without reading the article. Participants had 4 attempts to answer the questions correctly for the article.} In the case of 4 wrong answers, participants would be logged out of the session and considered unsuitable for the task. \rev{No participants in the study were disqualified through this method.}
     \item \textit{Ask two or more questions relevant to the article:}
      We asked participants to provide at least two questions relevant to the article. Participants could ask extra questions if they want. We prompted participants to ask about clarifications, follow-up information, 
	or any other questions relevant to the article. Depending on the condition (A or C), we presented the scenario of asking the questions to the authors or a chatbot. Participants had access to the PDF in this step. \rev{We alerted participants again that the research team will manually validate each question and any suspected fraud or abuse will result in forfeiture of compensation and removal of data.} 
    \item \textit{Rate the quality of articles and provide expertise:} 
    To answer RQ3, we wanted participants to rate the quality of the articles. However, quality is a complex construct, combined with many factors~\cite{doi:10.1177/107769909907600213}. Graefe et al.~\cite{doi:10.1177/1464884916641269} used credibility, readability, and journalistic expertise to measure quality. These three dimensions are calculated from 16 factors (e.g., accurate, trustworthy, entertaining, coherent, etc.). We determined that 16 factors would be too many for participants to rate. Thus, we asked participants to rate the credibility, readability, and journalistic expertise of the article on a scale from 1 to 5 (higher is better). We asked participants to consider the 16 contributing factors while rating the 3 higher-level dimensions. 
    % Participants rated each article on a scale of 1 to 5 (higher is better) across three dimensions: Readability, Credibility, and Journalistic Expertise~\cite{doi:10.1177/1464884916641269}.
    Finally, participants self-reported their knowledge of the broader topic of the article on a 4-point scale (1= not at all, 4= very knowledgeable).
 \end{enumerate}

 \item \textbf{Complete demographic survey:} Participants completed a demographic survey and received a unique code to indicate the completion of the study in MTurk.
 \end{enumerate}

% \rev{We determined this protocol by conducting a pilot study with 10 undergraduate and graduate students at our university.  Overall, participants found the process of reading 3 articles to be easy and quick. This also allowed us to assign one article from each domain to each participant and curate a wide range of questions. }

% We varied the number of articles to read (1-4) between different participants.

% (e.g., how many articles a participant should read? and how much time a participant should spend in the study?),

% \subsection{Quality Control}

% \renewcommand{\arraystretch}{0.8}

\subsection{Quality Control}
We ran the study in batches, 10 users at a time. After each batch, we manually examined each response to evaluate the quality of the questions. We discarded 2 responses (participants) due to incomplete submissions. \rev{This indicates that 99\% participants were able to successfully complete the study without compromising the quality of the responses.}

\subsection{Analysis}
% We analyzed the questions using both automated and manual labels across 14 dimensions (Table~\ref{tab:codebook}).

We used Grounded Theory~\cite{glaser2017discovery} to analyze the questions. All manual coding was conducted by two authors. At first, we selected 20\% of the questions randomly and two authors open-coded this subset independently. The authors then met to finalize the codebook, discussing different dimensions and observations.  
The authors then coded this set again independently by following the codebook. The inter-annotator agreement after this stage was 0.80 (Jaccard's Similarity). The disagreements at this point were resolved through discussion between the two coders, with input from the full research team. The rest of the questions were divided equally between the coders. Throughout this process, the authors met and discussed regularly about the codes. 

Beyond grounded theory, we coded the complexity of each question based on Bloom's Taxonomy. \rev{We followed a similar protocol as above for coding this dimension. The inter-annotator agreement was 0.82 (Jaccard's Similarity) for this dimension.} We also coded the questions based on whether they indicate a violation of journalistic integrity (Truth and Accuracy, Independence, Fairness and Impartiality, Humanity, and Accountability), according to the Ethical Journalism Network~\cite{ejn}. \rev{The inter-annotator agreement was 0.92 (Jaccard's Similarity). The disagreements for both dimensions were resolved through discussion.}
Finally, we also conducted several automated analyses (e.g., sentiment and readability index) on the questions. We provided details of each automated analysis in the relevant sections.

% We coded the questions based on the five core principals (Truth and Accuracy, Fairness and Impartiality, etc.) from Ethical Journalism Network~\cite{ejn}.

% we need to clarify that all coding no matter bloom, or integrity or taxonomy were done by multiple coders. In the 4.7 we need to clarify that we went beyond grounded theory. For instance, we can mention there briefly about measuring complexity with bloom and finding ethical concerns using the Ethical JOUr Network

% We will group codes to higher level concepts throughout this process.

% condition	index	authors	chatbot
% 0	Fact	75	136
% 1	Details	174	128
% 2	Evidence	36	20
% 3	Close_ended	25	25
% 4	Opinion	51	18
% 5	Explanation	114	83
% 6	Other	2	7

\begin{figure*}
    \centering
    \includegraphics[width=0.7\textwidth]{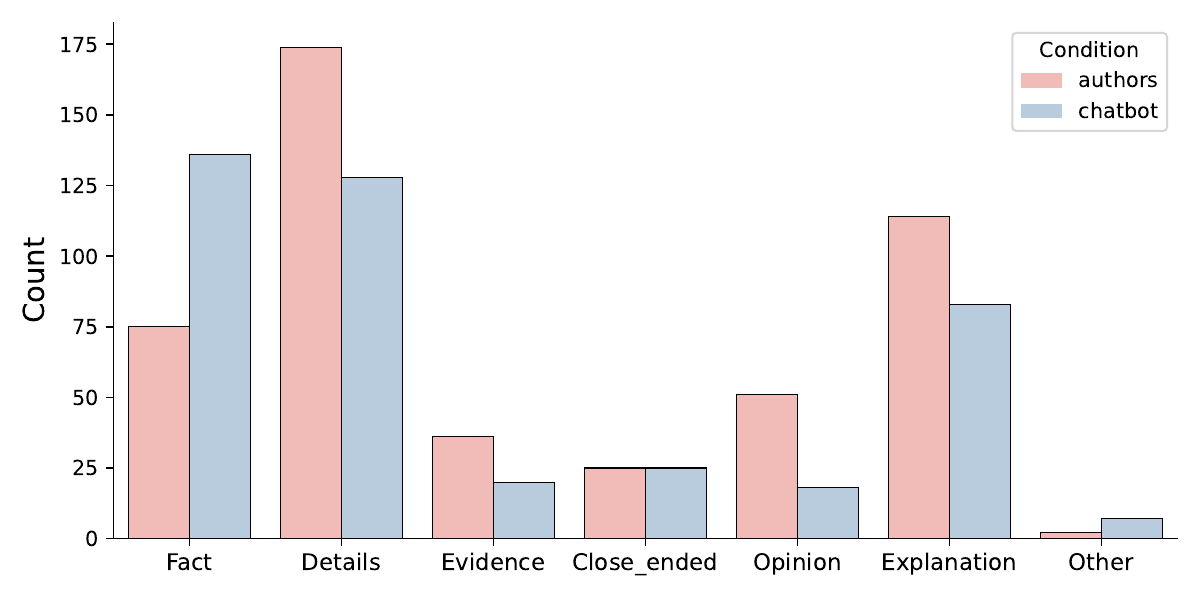}
    \caption{\textbf{Number of questions asked to the authors and chatbot.}}
    \label{fig:taxonomy_dist}
    \Description{This figure shows a bar chart with multi-colored bars.}
\end{figure*}

\section{Results from the Study with Audience}

We collected \rev{752 questions (3 articles * 2 questions * 124 participants = 744 + 8 extra) from MTurk}. This section presents the results of our study. We structure the results around our RQs.

\subsection{A Taxonomy of Questions (RQ1)}
\label{sec:rq1}
Table~\ref{tab:types} shows different types of questions we identified from the study.  We found three high-level categories: Information, Interpretation, and Others. The high-level categories are deconstructed into seven sub-categories. Among them, three categories (Factual, Opinion, and Explanation) align with the taxonomy identified by prior research on news comment sections~\cite{DBLP:conf/icwsm/RischK20, DBLP:conf/cikm/RischK20}, while others are new. Unlike previous taxonomies, we found that the categories \rev{are not mutually exclusive; a question can fall under multiple categories.} For example, consider the following question: 

\begin{quote}
    \textit{\rev{How would you compare the level of energy and excitement for Trump at this event to that of his announcement for his presidential bid for 2016?}}
\end{quote}

\rev{This question is expecting more \textit{details} about a current event and a previous event from 2016. It also seeks to know the author's \textit{opinion} about the energy and excitement levels in the two events. Similarly, this question, \textit{``What are the Northern Triangle countries and what makes them so important?''} expects both \textit{factual} information about the Northern Triangle countries and an \textit{explanation} about why they are so important. \rev{83.24\% (626/752)} questions belonged to only one category, with \rev{14.63\% (110/752)} and \rev{2.13\% (16/752)} questions belonging to two and three categories, respectively. }

% \begin{figure}
%     \centering
%     \includegraphics[width=0.95\textwidth]{figures/reference.pdf}
%     \caption{\textbf{References in the questions.} (left) Number of questions in reference to the title and first paragraph (i.e., summary) of the articles. (middle) Number of questions in reference to the main body of the articles. (right) Number of  second person reference (you, yours, etc.) in the questions.}
%     \label{fig:enter-label}
% \end{figure}

% 	index	Condition	Value
% 0	Fact	authors	75
% 1	Details	authors	174
% 2	Evidence	authors	36
% 3	Close_ended	authors	25
% 4	Opinion	authors	51
% 5	Explanation	authors	114
% 6	Other	authors	2
% 7	Fact	chatbot	136
% 8	Details	chatbot	128
% 9	Evidence	chatbot	20
% 10	Close_ended	chatbot	25
% 11	Opinion	chatbot	18
% 12	Explanation	chatbot	83
% 13	Other	chatbot	7

\begin{figure*}
    \centering
    \subfloat[]{
        \includegraphics[width=0.58\textwidth]{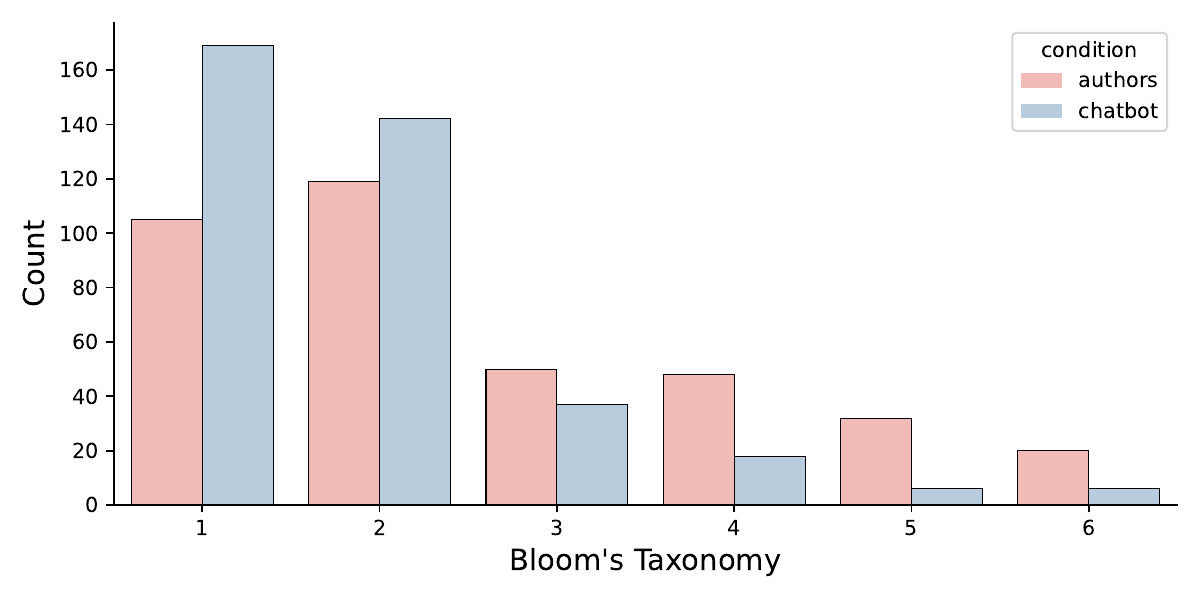}
    } 
    \subfloat[]{
        \includegraphics[width=0.4\textwidth]{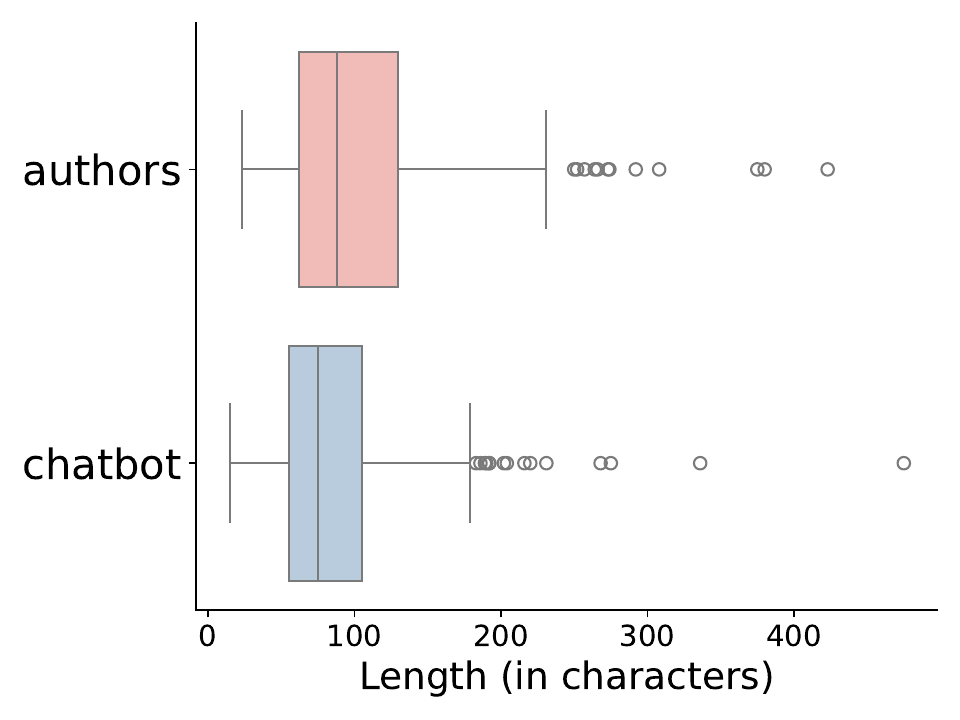}
    }
    
    \caption{\textbf{Complexity of the questions}. (a) Rating of the questions based on Bloom's Taxonomy. Participants asked questions higher on the taxonomy more frequently to the authors than the chatbot. (b) Length of the questions. On average, questions asked to the authors had higher length than those asked to the chatbot. }
    \label{fig:complexity}
    \Description{This figure has two sub-figures. The left sub-figure shows a bar chart with colored bars, and the right sub-figure shows a box plot with colored boxes.}
\end{figure*}

\begin{figure*}
\centering
    \subfloat[]{
        \includegraphics[width=0.35\textwidth]{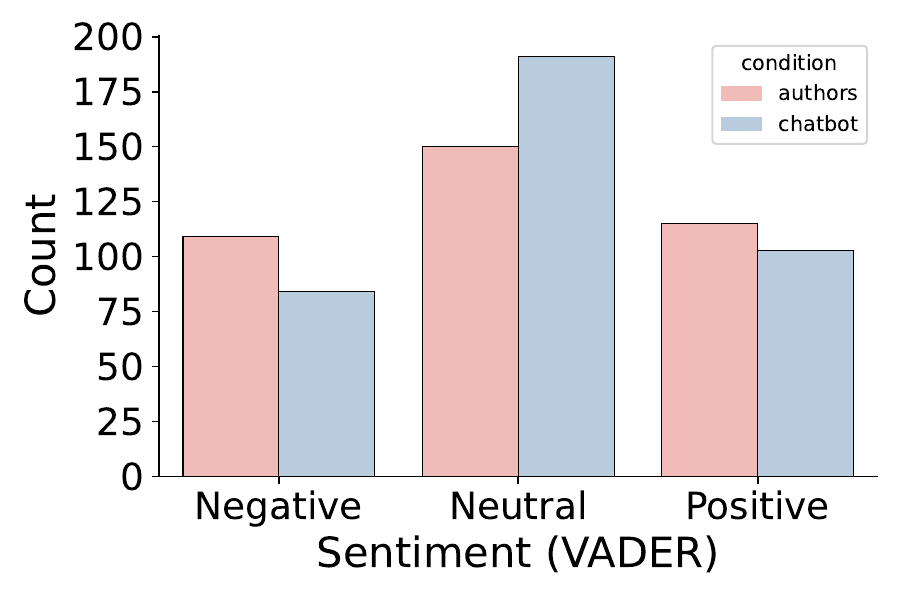}
        \label{fig:sentiment}
    }
    \subfloat[]{
        \includegraphics[width=0.28\textwidth]{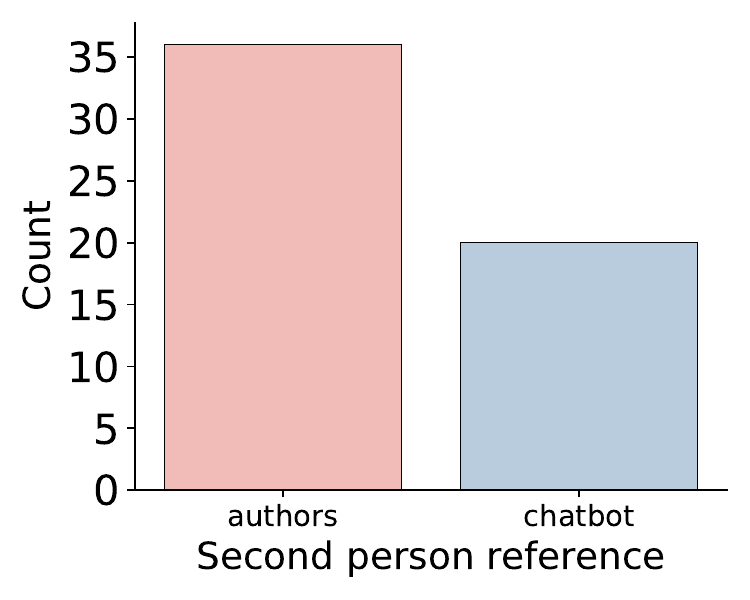}
        \label{fig:second_person}
    }
    \subfloat[]{
        \includegraphics[width=0.35\textwidth]{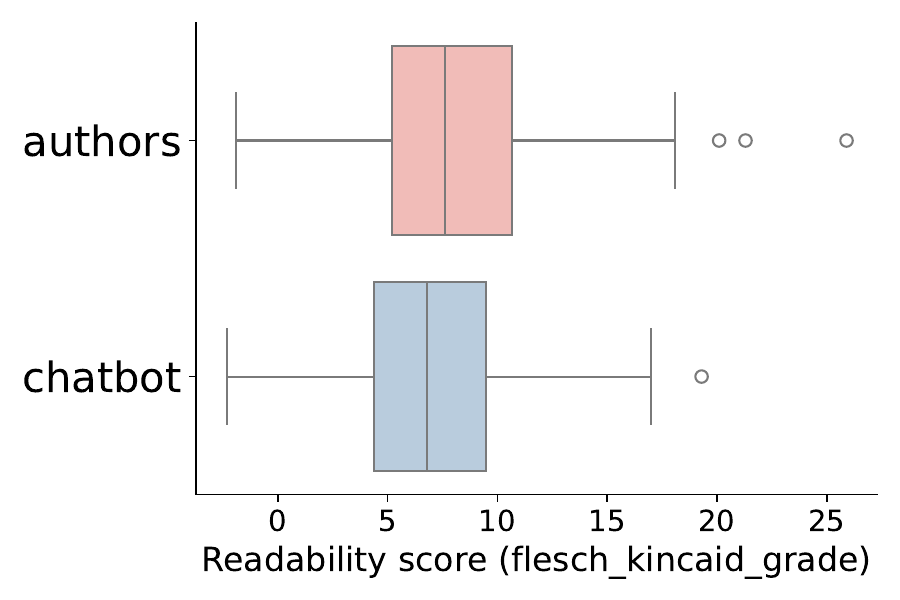}
        \label{fig:readability}
    }\hfill
    \caption{\textbf{Linguistic features of the questions}. (a) Sentiment of the questions based on the VADER model~\cite{DBLP:conf/icwsm/HuttoG14}. (b) Second person (you, yours, etc.) reference in the questions. (c) Readability Index (Flesch Kincaid Grade).}
    \label{fig:linguistic}
    \Description{This figure has three sub-figures. The left and middle sub-figure show bar charts with colored bars, and the right sub-figure shows a box plot with colored boxes.}
\end{figure*}

\begin{figure*}
    \centering
    \includegraphics[width=0.8\textwidth]{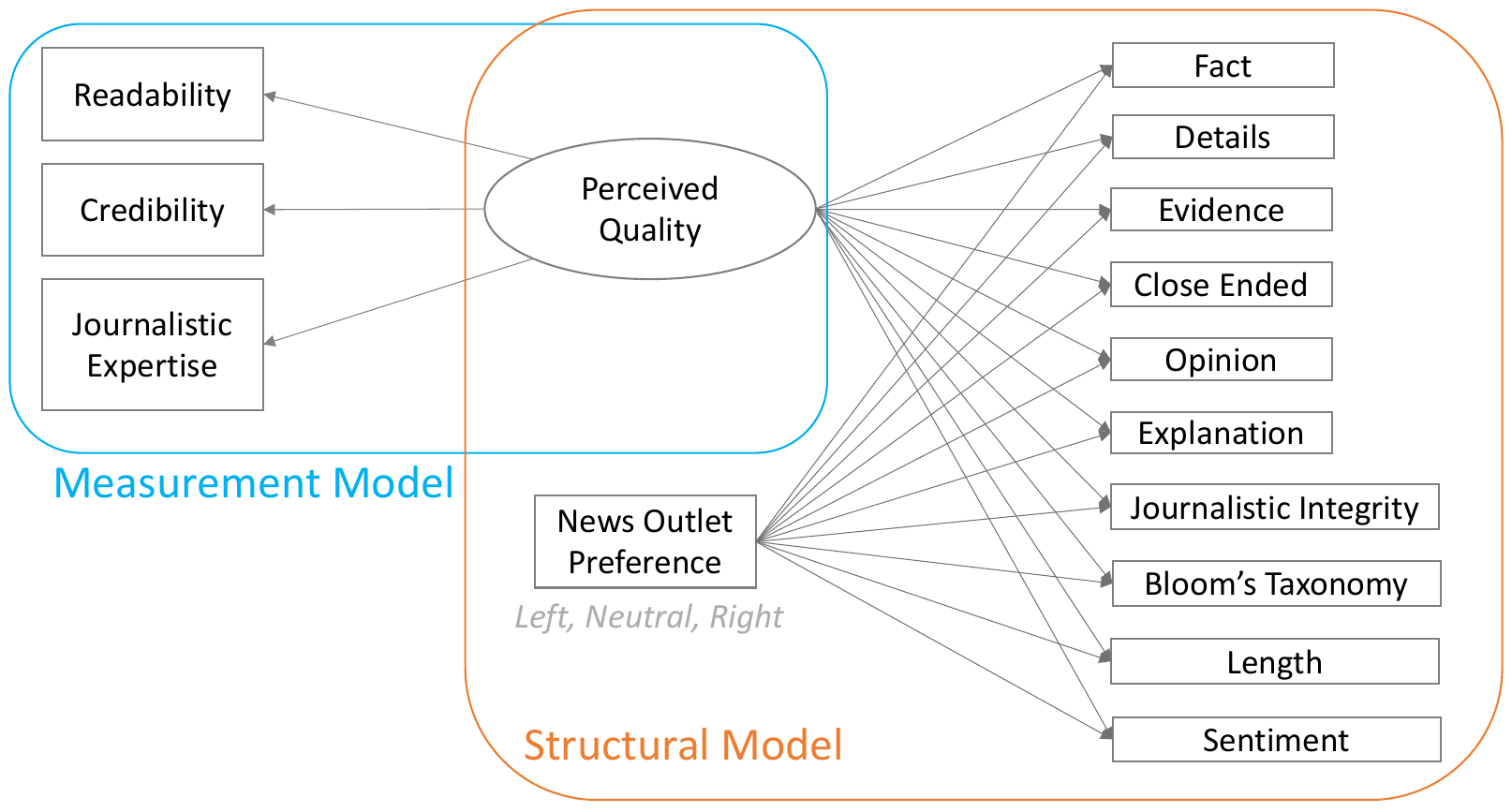}
    \caption{\textbf{Structural Equation Model (SEM) for measuring the effect of perceived quality and news outlet preference.} Here, perceived quality is a latent variable, derived from readability, credibility, and expertise ratings provided by the participants. We determined participants' preference for news consumption (left, neutral, right) from their reading preference (Figure~\ref{fig:participant}e).}
    \label{fig:sem_model}
    \Description{This figure shows multiple visual elements and mapping between them.}
\end{figure*}

\subsection{Authors vs Chatbot (RQ2)}

\subsubsection{Taxonomy Distribution}
Figure~\ref{fig:taxonomy_dist} shows different types of questions that were asked to the authors (A) and chatbot (C) by the participants. Participants asked significantly more factual questions to the chatbot compared to the authors \rev{(A= 75, C= 136)}. In comparison, participants asked more questions seeking \rev{long-form details (A=174, C=128), evidence (A=36, C=20), opinion (A=51, C=18), and explanation (A=114, C=83) to authors}. Finally, participants asked a similar amount of \rev{close-ended questions (A=25, C=25)} and a small number of questions that did not fall into any of the categories \rev{(A=2, C=7)} using both conditions. 

% \rev{We also calculated how participants routed questions between the two high-level categories of the taxonomy. 17 out of 62 (27.4\%) participants for the condition \textbf{C} asked questions seeking only information (i.e., all 6 questions from these participants sought information). 45 (72.6\%) participants had questions that fall under both Information and Interpretation. }

\subsubsection{Complexity}
The distribution in taxonomy already shows that participants asked questions that require subjective thought, opinion, and knowledge more frequently to the authors than the chatbot. We further labeled the questions based on  Bloom's taxonomy~\cite{krathwohl2002revision} to measure the complexity and depth of knowledge required to answer the questions. Bloom's taxonomy is a six-point hierarchical scale (1= Knowledge, 6= Evaluation) where higher levels indicate a higher knowledge, learning, and cognitive skills requirement to answer the questions than previous levels.

Figure~\ref{fig:complexity}a shows the distribution of the questions based on Bloom's taxonomy. Given that most questions in our data seek facts and details, it was expected that levels 1 and 2 would be the most common according to Bloom's taxonomy. However, on average, participants asked more complex questions to authors \rev{($2.58$) than the chatbot ($1.86$)}. The Mann–Whitney U test showed that the difference was statistically significant ($p< 0.0001$). We further computed the length of the questions \rev{($Mean_A= 102.90, Mean_C= 86.26$)}. The difference was statistically significant with $p< 0.007$ (Figure~\ref{fig:complexity}b). Finally, we counted questions that fall under 2 to 3 categories in our taxonomy at the same time (Section~\ref{sec:rq1}) as that is a measure of questions having multiple facets. Out of \rev{110} questions that had two categories or facets, readers asked \rev{71 to the authors (64.55\%)}. All \rev{16 questions} that belonged to 3 categories were asked to the authors.
% Readers asked all 15 questions that had three categories to the authors.

% Overall, fact and details seeking questions are the two most common categories. However, participants asked for short and factual information more frequently to chatbot than authors (A= 63, C= 113) whereas the trend was opposite for long form details seeking questions (A=141, C=100).

% \begin{figure}
%     \centering
%     \includegraphics[width=0.9\textwidth]{figures/automated.pdf}
%     \caption{\textbf{Features calculated from the questions automatically.} (top left) Distribution for the length of the questions; (top right) Distribution for the readability score of the questions  based on Flesch Kincaid Grade; (bottom left) Sentiment score of the questions using VADER~\cite{DBLP:conf/icwsm/HuttoG14}; and (bottom right) Sentiment score of the questions using Textblob.}
%     \label{fig:automated}
% \end{figure}

\begin{table*}[]
    \centering
    \begin{tabular}{m{1.7cm}|m{1.5cm}m{1.2cm}m{1.2cm}m{1.2cm}|m{1.5cm}m{1.2cm}m{1.2cm}m{1.2cm}}
    \toprule
     & \multicolumn{4}{c}{\textbf{Authors}} & \multicolumn{4}{c}{\textbf{Chatbot}} \\
    \hline
    & \textbf{Perceived Quality}  & \textbf{Left Alignment} & \textbf{Neutral Alignment} & \textbf{Right Alignment} & \textbf{Perceived Quality}  & \textbf{Left Alignment} & \textbf{Neutral Alignment} & \textbf{Right Alignment} \\ 
    \hline
    \textbf{Fact} & -0.13* & 0.04 & -0.02 & -0.03 & -0.14 & 0.04 & -0.02 & -0.03 \\
    \hline
    \textbf{Details} & 0.42*** & -0.07 & -0.09 & 0.15 & 0.35** & -0.07 & -0.09 & 0.15 \\
    \hline
    \textbf{Evidence} & -0.40*** & 0.05 & 0.03 & -0.09 & 0.10 & 0.05 & 0.03 & -0.096 \\
    \hline
    \textbf{Closed-ended} & 0.003 & -0.12 & 0.16 & 0.25 & 0.12* & -0.12 & 0.16 & 0.25 \\
    \hline
    \textbf{Opinion} & 0.22* & 0.01 & 0.05 & -0.07 & 0.04 & 0.01 & 0.05 & -0.07 \\
    \hline
    \textbf{Explanation} & -0.27** & 0.04 & 0.08 & -0.08 & 0.06 & 0.04 & 0.08 & -0.08 \\
    \hline
    \textbf{Journalistic Integrity} & -0.54*** & 0.07 & 0.06 & -0.13 & -0.11 & 0.07 & 0.06 & -0.13 \\
    \hline
    \textbf{Bloom's Taxonomy} & 0.06 & 0.20 & -0.03 & -0.17 & 0.14 & 0.20 & -0.03 & -0.17 \\
    \hline
    \textbf{Length} & 14.70 & 24.22 & 3.60 & 30.04 & 11.23 & 24.22 & 3.60 & 30.04 \\
    \hline
    \textbf{Sentiment} & -0.03 & 0.007 & -0.03 & -0.001 & -0.05 & 0.007 & -0.03 & -0.001 \\
    \bottomrule
    \end{tabular}
    \caption{\rev{\textbf{Results of the SEM model.} Each cell represents the coefficient for the respective relation. \revn{Same coefficient values between the author and chatbot conditions indicate globally constrained paths in the SEM model (i.e., no differences between the groups) based on the multigroup analysis}. p-value significance: * p < 0.05, ** p < 0.01, *** p < 0.001. }}
    \label{tab:report}
    \Description{A table with ten rows and two columns. Both column has four sub-columns.}

\end{table*}

\subsubsection{Questioning journalistic integrity}
We noticed that a few questions focused on the integrity and ethical dimensions of reporting. For instance, consider the following two questions- both indicating a violation of Truth and Accuracy~\cite{ejn}.  The first one is directed toward the authors of an article while the second one is toward the chatbot:

\begin{quote}
    ``Why would the author imagine this has something to do with the veracity of what’s being said on the air when the entire article only details allegations that the people referenced engaged in inappropriate conduct behind the scenes?''

    ``Are you sure that low-income people were at higher risk of depression, they might have already had depression before that. How was it tested to get that result?''
\end{quote}

 In total, we found \rev{73 ($9.71\%$)} questions raised doubts related to at least one of the five principles~\cite{ejn}. Out of the \rev{73 questions, 49 (67.12\%)} were directed toward the authors, while \rev{24 (32.88\%)} were directed toward the chatbot.

\subsubsection{Linguistic Features}
We measured several linguistic features of the questions. Figure~\ref{fig:sentiment} represents the frequency of different sentiment categories (Positive, Negative, Neutral) for questions directed toward authors and chatbots, based on the VADER sentiment analysis model~\cite{DBLP:conf/icwsm/HuttoG14}. The figure suggests readers had more emotionally charged questions for authors, especially with negative connotations, than chatbots. The reason may be because readers asked more critical questions (e.g., about journalistic integrity) to the authors. Readers directed more neutral questions toward chatbots than authors. The reason may be because most questions asked to chatbots seek simple facts and details without any particular sentiment.

A second-person reference (e.g., you, yours, etc.) in a question means that the question is directed at the authors or chatbot. Such references often create a more personal and engaging tone, as they involve the reader of the question directly in the topic being discussed~\cite{doi:10.1177/0261927X09351676}. We found that participants used such references more frequently to address the authors than chatbot (Figure~\ref{fig:second_person}).

Finally, we found that questions asked to the authors have a higher readability index according to the Flesch Kincaid Grade scale (Figure~\ref{fig:readability}). Although the effect size was small \rev{($\Delta = 0.89$)}, the difference was significant with $p<0.01$.

% sentiment most pos or neg?
% We also measured sentiment scores (VADER and TextBlob) and Named Entity Recognition (NER). However, we did not find any significant differences. 

% \subsubsection{Automated Analysis}

\subsection{Effect of Perceived Quality and Personal Preferences (RQ3, RQ4)}
\label{sec:sem}

To measure how perceived quality influences questions, we adopted a structural equation model (SEM)~\cite{ullman2012structural}, shown in Figure~\ref{fig:sem_model}. Similar to previous research~\cite{doi:10.1177/1464884916641269}, we consider perceived quality to be a latent variable and derive it from user-provided ratings on readability, credibility, and journalistic expertise. Figure~\ref{fig:quality} in Appendix A shows participants' ratings across the three higher-level dimensions that capture the perceived quality of the articles. Overall, participants rated the articles highly across the three dimensions. To answer RQ4, we also added participants' preferences for news outlets in the model. We manually checked the preferred news outlets provided by the participants (Figure~\ref{fig:participant}e), matched the polarity of the outlets to AllSides ratings~\cite{allsides}, and then categorized them to either left, neutral, or right-leaning news outlets. We assigned neutral to participants who read both left and right-leaning news \rev{(27)}. We assigned a left-leaning tag if the participant consumed mostly left-leaning news \rev{(60)} and right-leaning \rev{(37)} vice versa. \revn{Finally, we conducted multigroup analysis~\cite{jamesandjulia} to capture differences between the authors and chatbot condition. We used the \textit{piecewiseSEM}~\cite{piecewiseSEMpaper} package in \textsc{R} to find which paths vary across the two conditions (i.e., groups).}

Table~\ref{tab:report} presents the results from the SEM analysis. \revn{The model fit measures are: Comparative Fit Index (CFI): 0.96; Goodness of Fit (GFI): 0.90; Adjusted Goodness of Fit (AGFI): 0.91; $\chi^2$: 610.59; and RMSEA: 0.07. The measures indicate a good SEM model fit.}. The coefficients show the polarity and strength of the relations. For example, the coefficient for Journalistic Integrity and Perceived Quality was \revn{-0.54}. This means low-quality articles correlated with a high number of questions about journalistic integrity. 

When interacting with the authors, we found that high-quality articles encouraged participants to seek more \textit{details} and \textit{opinion} about the topic. Readers were more critical of authors when the perceived quality was low, seeking more \textit{facts} and \textit{evidence}, and questioning \textit{journalistic integrity}. In contrast, when interacting with the chatbot, readers were less critical, even if the perceived quality was low. Thus, it implies that readers may not engage with a chatbot even if they think the quality of the article is poor, and many critical questions can be raised. We did not observe any significant impact of readers' choice of news outlets in the questions.

\begin{figure}[tbh]
    \centering
    \includegraphics[width=\columnwidth]{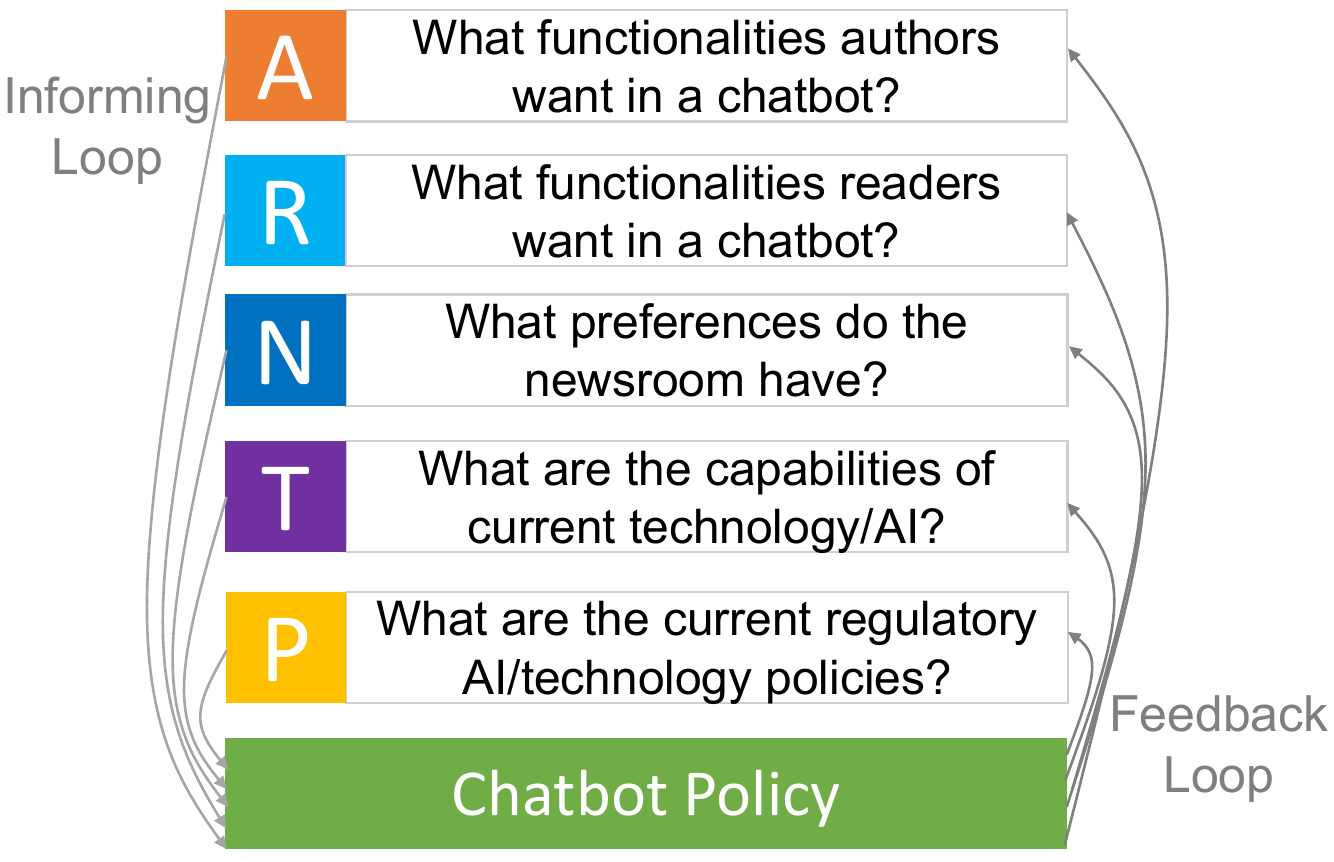}
    \caption{\textbf{Framework for designing chatbots in online news.} Our framework includes \revn{five} different facets informing the policy. Once the policy is formalized, it needs to be communicated through the feedback loop. The informing and feedback loop can be iterated multiple times. }
    \label{fig:framework}
    \Description{This figure shows multiple text boxes and mapping between them.}
\end{figure}

\section{A Framework for Designing Chatbots}

\rev{
Our findings from the study with the audience indicate that readers' questioning pattern varies significantly based on the receiver of the questions---authors vs. chatbot. We noticed this difference across various dimensions---from the type, complexity, and linguistic features of questions to the user behavior towards journalistic integrity and perceived quality. The logical conclusion is that readers perceive authors and chatbots (or machines in general) very differently. Readers' lack of trust in AI and skepticism about AI are likely the reasons behind this difference~\cite{DBLP:conf/fat/KnowlesR21, 10.1145/3531146.3533077}. 
Increased scrutiny in popular media and legal steps from different domains such as the introduction of the ``Blueprint For An AI Bill Of Rights'' from the White House~\cite{blueprint} may have further contributed to such polarized  view. 

While infrequent, we found evidence that readers may project human-like feelings and behaviors onto AI (i.e., anthropomorphizing AI~\cite{Deshpande2023}). For instance, a considerable amount of readers want the chatbot to be able to provide answers to subjective questions, albeit not as frequently as to factual questions. Readers' prior experience with chatbots such as ChatGPT and narratives about Artificial General Intelligence (AGI) may have contributed to this perception~\cite{agi_times}. }

\rev{

Overall, the findings from the two studies suggest that there is a high-level agreement between journalists and readers about the role of a QA chatbot in newsrooms. Both parties predominantly wanted the chatbot to answer factual questions while they wanted to engage with each other directly for subjective questions. However, as mentioned above, our findings suggest readers do not see the role of \revn{such a chatbot} to be a \textit{zero-sum situation}, i.e., all factual questions should be answered by the chatbot while journalists should focus on answering questions requiring subjective interpretation. A considerable amount of readers wanted the chatbot to answer factual as well as subjective questions. Similarly, we also noticed that readers want authors to answer a substantial number of factual questions, despite asking most of the factual questions to the chatbot.}

% Similarly, }

% A considerable amount of participants still want authors to answer factual and detail-seeking questions. Similarly,

\rev{Given this somewhat conflicting alignment between readers and journalists, it is clear that researchers and news organizations need a systemic approach to calibrate and communicate the needs and expectations of the two stakeholders. We propose a conceptual framework with four facets to facilitate this. For ease of explanation, we assume that a newsroom or news organization is developing the QA chatbot. However, other user groups (researchers, tech industry) can use our framework as well. Finally, while our discussion below is specific to QA chatbots, we believe the framework will be useful for designing a wide range of chatbots. }

% Based on the findings from \rev{the study with journalists, readers, and prior literature}, we propose a framework for designing QA chatbots (Figure~\ref{fig:framework}). For ease of explanation, we will assume that a research group is developing the chatbot. However, other user groups (news organizations, tech industry) can use our framework as well. 

% \rev{Findings from the two studies suggest that there is a high-level agreement between journalists and readers about the role of a QA chatbot in newsrooms. Both parties predominantly want the chatbot to answer factual questions while both parties want to engage between themselves for subjective questions and conversations. However, our findings suggest we should not assume it be a \textit{zero-sum situation} and }

% and two phases: \textit{Inform} and \textit{Feedback} loop. We describe each phase below.

% \subsubsection{Inform Loop}
% Our framework includes four different facets. 
\subsection{Facets}
We briefly describe the \revn{five} main facets or components of the framework below.

\textbf{Authors.}
\rev{The first facet of the framework is the Authors.} Since authors or journalists are the ones that are typically responsible for answering questions or response to readers, it is essential that the chatbot serves authors' needs and does not overstep into authors creative space. News organizations may conduct interviews such as ours with journalists to gather such information.

% News organizations should try to understand how authors or journalists want to be involved in the process.

\textbf{Readers.}
The second facet is understanding readers' needs. The central question for this facet is \textit{``What functionalities readers want in a chatbot?''} \rev{It is possible to design the chatbot just by collecting requirements from the journalists, but such a chatbot may not match readers' needs and improve audience engagement, which is the primary reason behind supporting question-answering~\cite{meier2018audience, nelson2021next}. }

% For instance, our findings suggest, similar to authors, readers too predominantly want the chatbot to answer factual questions while want to engage with authors for subjective questions and conversations. However, our findings suggest we should not assume it to be a \textit{zero-sum situation}, that chatbots should only answer factual questions and route all other questions to the authors. Readers expect the chatbot to answer questions requiring subjective thoughts. Similarly, readers do not want authors to completely abandon answering factual questions. We describe ways to match this somewhat paired views between journalists and readers below.}
% To answer this question, news organizations may conduct medium to large-scale online experiments such as ours. Alternatively, news organizations may conduct one-to-one interviews, or arrange focus groups to answer the question.

\textbf{News Organization.}
The third facet is the preferences of the news organization itself. \textit{Should each article have a chatbot option? Does the chatbot need to adapt to the domain of the article (e.g., health) and types of reporting (e.g., investigative vs. beat reporting)?} These questions need to be answered in this facet.

\textbf{Technology.}
The \revn{fourth} facet is understanding the design space of the currently available technology.  News organizations should conduct a review such as ours in Section 2, consult internal research and developer teams, or recruit external researchers or corporate tech companies for this purpose.

\revn{\textbf{AI Regulatory Policies.}
The final facet is seeking legal consults to understand current AI/technology regulatory policies such as the  ``General Data Protection Regulation (GDPR)''~\cite{voigt2017eu} from the European Union (EU) or the ``Blueprint For An AI Bill Of Rights''~\cite{blueprint} from the White House in the U.S. These policies may contain rules and policies that news organizations should abide by while designing the chatbot.}

\subsection{Inform Stage}
The four facets should inform the chatbot policy. Researchers should balance requirements from the facets at this stage. Here, based on the findings from our studies, we demonstrate \textit{three examples} on  how researchers can balance requirements at this stage.

\subsubsection{Developing Chatbots for Answering Factual Question.} 
\rev{The two arrows from the authors and readers in Figure~\ref{fig:framework} (through the two studies) have informed us that any future QA chatbot should be able to answer factual questions. However, the arrow from technology indicates that the most powerful technology available today, LLMs, can hallucinate, provide information without proper references, or even provide wrong information~\cite{DBLP:journals/corr/abs-2112-04359}. Thus, newsrooms should consider how they can support factual QA during this stage. For instance, instead of relying solely on the LLMs, researchers can consider a hybrid model that uses an expert user base for this purpose. Xiao et al. recently showed that coupling AI with an expert user base can support the health information needs of general users~\cite{DBLP:conf/iui/XiaoLZG023}. Researchers from Microsoft Research recently proposed LLM-Augmenter~\cite{DBLP:journals/corr/abs-2302-12813}, a system that can provide answers to factual questions by using external knowledge bases. One good knowledge base for news articles could be the background research and references collected by the authors for writing the articles~\cite{caswell2019structured, caswell2018automated}.}

% Also, answering factual questions would require journalists and other personnel involved in the news production to follow structured journalism~\cite{caswell2019structured, caswell2018automated} guidelines so that all the background reference information are searchable.}

\subsubsection{Routing Factual and Subjective Questions.}
\rev{Our findings indicate that researchers will need to devise a routing policy for factual and subjective questions in the future. One practical solution would be to adopt the predominant view between journalists and readers: the chatbot will only answer factual questions while re-routing all questions seeking opinions and explanations to the authors. While this policy may not cater to the requirements of every reader, it will likely guarantee legal compliance for the news organization. For example, \revn{the AI regulatory policy facet in the framework can inform the newsroom that asking a chatbot to provide human-like opinion may violate  legislative principles penned in the ``Blueprint For An AI Bill Of Rights''~\cite{blueprint, Deshpande2023}.}}

\rev{Newsrooms can adopt machine learning approaches to achieve this routing mechanism. For example, newsrooms can use soft-prompting~\cite{qin-eisner-2021-learning} with a pre-trained LLM to separate factual questions from subjective questions. Alternatively, newsrooms can collect a large corpus of questions similar to ours and then fine-tune a pre-trained LLM for this task. Finally, our findings suggest readers may expect answers to some factual questions directly from the authors, instead of the chatbot. To facilitate that, newsrooms can develop a scoring model that will measure the severity of a factual question and based on that route the question to either the authors or chatbot. }

\subsubsection{Collecting Critical Feedback from Readers}

\rev{Our findings suggest \revn{that} over-reliance on chatbots may deprive newsrooms of obtaining critical feedback from readers such as the underlying causes for the low perceived quality of an article or what raised questions about journalistic integrity in readers' minds. According to our results, readers may not raise questions about journalistic integrity to a chatbot as frequently as they would to the authors, even if they think the quality of the article is low. Several previous works have reported similar negative \revn{impacts} of chatbots in user experience design~\cite{10.1145/2858036.2858288, 10.1145/2901790.2901842}.  Considering this fact, researchers may recommend news organizations to plan alternative methods to collect critical feedback from readers. As mentioned by journalists in the interviews, many organizations now organize audience gathering and feedback sessions, which could be an excellent alternative to achieve this.}

% We believe this phenomenon is related to the concept of anthropomorphizing AI~\cite{Deshpande2023}. While many  . 

% We can not say for sure which reason driving the situation, but more focused  }

% For example, \rev{to balance the requirements from journalists and readers, researchers can collect a large corpus of questions similar to ours and then built a predictive model that will be able route }

% our findings suggest readers mostly want to ask questions seeking facts and details. However, the most powerful technology available, LLMs, can hallucinate, provide information without proper references, or even provide wrong information~\cite{DBLP:journals/corr/abs-2112-04359}. Thus, they are not suitable for answering facts and details yet. News organizations could consider a hybrid model that uses an expert user base for this purpose~\cite{DBLP:conf/iui/XiaoLZG023}. 

% Alternatively, news organization can decide not to support answering facts and details.

\subsection{Feedback Stage}
The final part of our framework is the feedback loop. All stakeholders from the \revn{five} facets should know the policy determined after the inform stage. \rev{For example, based on the discussion in Section 6.2.2., if  the news organization decides that the chatbot will not provide opinions and explanations, that decision should be conveyed to the readers. To encourage readers to provide critical feedback, news organizations can inform readers that questions about journalistic integrity will be immediately rerouted to authors. News organizations should also advertise other mediums for providing feedback (e.g., member gatherings) to readers. Finally, \revn{the} news organization can iterate on the two stages for finalizing the chatbot policy.}  

% Our findings suggest that readers also seek explanations and opinions from the chatbot, albeit infrequently. 

\section{Discussion, Limitation, and Future Work}

In this section, we discuss design implications, limitations, and future improvements of our work.

\subsection{Design Implication}
\subsubsection{Future of Audience Engagement}
    As we look towards the future, the landscape of audience engagement is poised for a transformative shift. With the integration of LLMs, readers could have interactive platforms catering to their immediate factual queries while valuing human authors for in-depth analyses and opinions. This symbiotic relationship between AI and human expertise can redefine news consumption, making it more interactive, personalized, and engaging. \rev{We have presented several design prospects and implications for future AI and HCI research in Section 6. } 
    % With the integration of LLMs, there is potential for a more personalized news experience where users could receive content tailored to their interests, previous interactions, and queries. Such personalization, combined with the interactivity of LLMs, can redefine user engagement metrics. 
    % On the flip side, it is possible that readers might get frustrated by seeing lackluster and stereotyped text~\cite{DBLP:conf/chi/MirowskiMPE23, DBLP:conf/ACMdis/GeroLC22} outputs from LLMs. This could lead to news organizations completely abandoning this technology for audience engagement.

% \subsubsection{Advancing Computational Journalism}
%     Our findings show that readers tend to ask factual questions more to chatbot compared to authors. Also, many evidence seeking questions were asked. Novel computational techniques need to be developed to ensure that Generative AI can accurately present evidences with references. This is one of the limitations of many current LLMs (e.g., ChatGPT 3.5). Also, answering the questions would require journalists and other personnel involved in the news production to follow structured journalism~\cite{caswell2019structured, caswell2018automated} guidelines so that all the background reference information are searchable.

\subsubsection{Increase Trust and Accountability}
    News organizations are grappling with a crisis of trust among their audiences. Prior research has \revn{demonstrated that} presenting information alongside explanations of how and why news stories were developed can significantly enhance the trust readers place in news outlets~\cite{masullo2017building}. This study adopts a novel approach by directly engaging with readers and inviting them to share their inquiries about the news. Successfully answering these questions about the news production process would quench \revn{readers' thirst for} news production issues (e.g., elaborating on the rationale behind including certain sources while omitting others) and would help increase trust and accountability.

\subsubsection{New Revenue Stream}
    The news industry, particularly local news is suffering from financial troubles~\cite{holyk2011news, kirchhoff2010us}. A new \revn{LLM-based} audience engagement could open a new revenue stream. For instance, the news industry can offer LLM-based chatbot services through API to third-party applications such as Google Home, Alexa, Siri, etc. Also, readers' questions can be utilized, with permission, for data mining and \revn{extracting} insights for better news \revn{recommendations and advertisements}. Moreover, each newsroom has its own archive of historical news. Currently, the archives are mostly reserved for record purposes without being involved in any revenue process. LLMs can utilize this valuable news archive to answer questions and bring this resource into a revenue process.   

% ---
% In summary, the findings from our study not only shed light on the current state of reader interactions but also provide a blueprint for the future of news engagement. By harnessing the strengths of both human authors and LLMs, we can envision a future where news consumption is a dynamic, interactive, and enriching experience.

\subsection{Limitations and Future Work}
Although we interviewed journalists and conducted an in-depth analysis of the readers' questions, our analysis is not free from limitations. 
For example, \rev{while we engaged a significant number of readers in the study, a small-scale followup interview study with readers might reveal readers' perspective about chatbots more clearly. We consider such a study as our immediate next step.} Our dataset contains several other facets that we have not explored yet. For example, we have not looked into the effects of news topics. We also have not looked into types of news such as investigative, breaking, etc. These analyses are left for future work as they were less relevant to our current research questions.

% We also did not study how the questions might differ between authors and non-authors or humans and AI. These are interesting facets that we want to study in the future.

\rev{The chatbot in our study did not feature any response or conversation, which likely would influence users' perception of AI and the chatbot~\cite{DBLP:conf/chi/LiaoSWV23}. Our future work will focus on designing an actual chatbot, informed by the findings of this paper, and then conducting a study to understand readers' questioning patterns while using the chatbot.}

\revn{The study might also involve hidden confounding variables.  For instance, there are two differences between the author and chatbot conditions: human vs AI and authors vs non-authors. We did not study these facets separately as our goal was to understand how readers direct questions towards authors and chatbot, not how the questions differ between human vs AI or authors vs non-authors. In other words, we were interested in the complete constructs embodied by authors and chatbot, not their individual components. Nevertheless, we acknowledge that studying these facets could provide more insights into readers' questioning patterns. } 

% Additionally, assuming AI as authors could lead to ethical problems associated with anthropomorphizing AI~\cite{Deshpande2023}; AI models are not persons and thus cannot be authors in the true sense, and there are legal, safety, security, trust, and reliability concerns in such relationships~\cite{Deshpande2023, Epstein_2023}. }

Finally, we have not conducted any experiment to evaluate how different LLMs such as GPT-4 and LLaMa might perform on our question set. We left this for future work as we believe that this itself is a separate research question. We aim to conduct the experiment with various QA systems and tasks: Closed domain vs. Open Domain~\cite{kwiatkowski-etal-2019-natural}, answerable vs. unanswerable~\cite{51002}, and so on.

% + We haven’t studied if the questions are indeed answerable
% + We need to look at quality of answers from LLMs

\section{Conclusion}
\revn{The} news industry is exploring how to integrate Large Language Models (LLMs) and Generative AI into news production, distribution, audience engagement, and other related processes. In this paper, we took the first step towards designing an LLM-powered chatbot by understanding how journalists and readers want to use such a chatbot. \revn{To achieve that, we interviewed journalists and conducted controlled experiments with readers.}  From interviewing journalists, we observe that journalists prefer a human-in-the-loop process rather than delegating all questions to a chatbot; particularly for the opinionated questions. We then examined what questions readers ask the authors and chatbot and if there are any differences. Our findings show that the questioning behavior of readers varies depending on who is receiving the question- the author or the chatbot. One of the major insights from our results is that readers ask factual questions more to chatbots compared to authors. Another significant observation is that readers question about journalistic integrity more to authors compared to the chatbot. Our findings have implications for designing a chatbot for audience engagement; specifically to evaluate how correctly LLMs can answer the questions that readers normally ask as we found. We argue that a careful and systematic approach to introducing this LLM technology to a trust-sensitive news industry would help minimize risks and ethical concerns. 

% \begin{acks}
% \end{acks}

\bibliographystyle{ACM-Reference-Format}
\bibliography{references}

\appendix

\section{Appendix A}
\begin{figure*}[h]
    \centering
    \includegraphics[width=\textwidth]{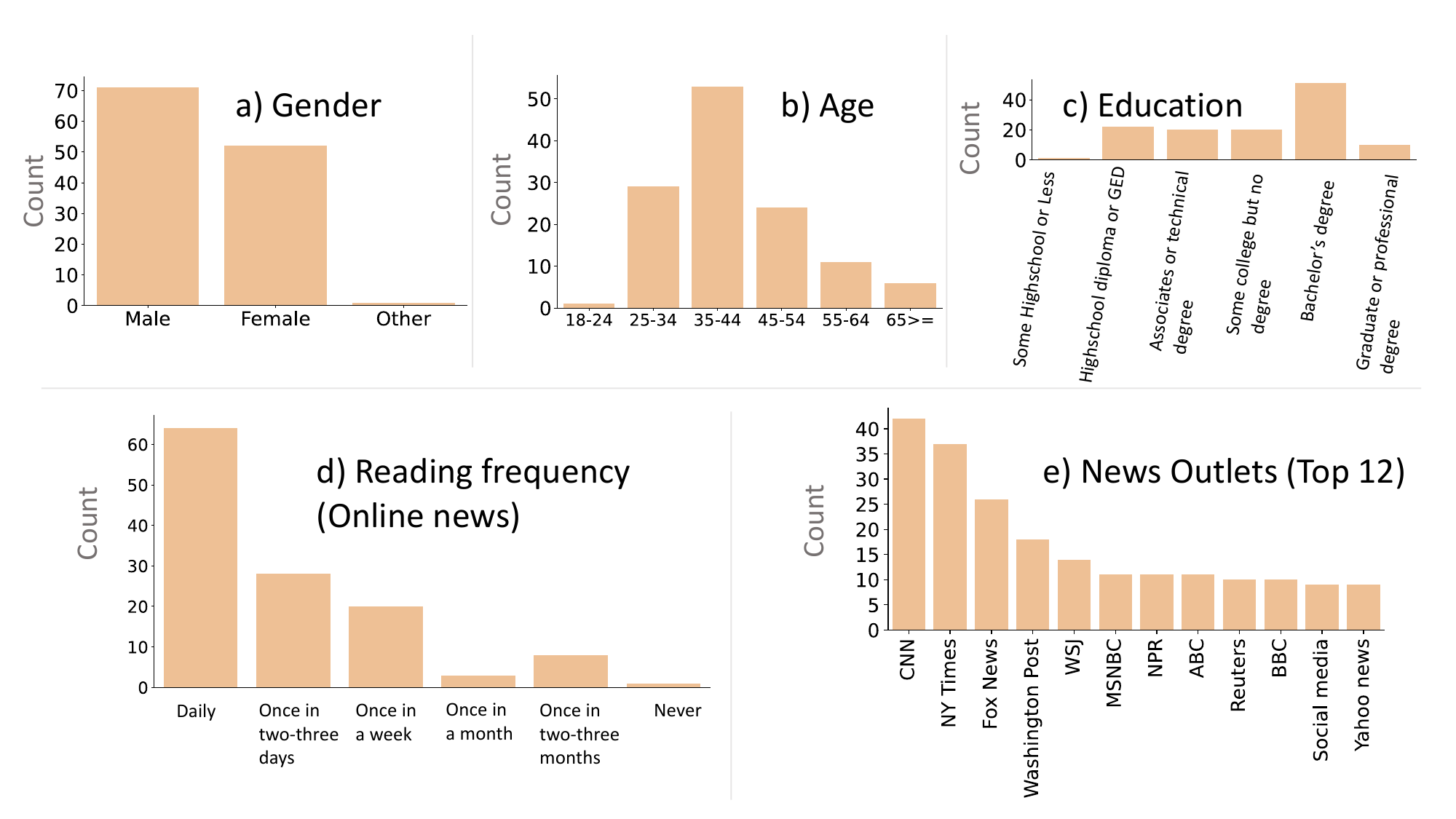}
    \caption{\textbf{Participant demographics.} Our participants (MTurks workers) varied in terms of a) gender, b) age, c) level of education, d) how frequently they read news online, and e) preferred news sources (top 12). }
    \Description{A figure with five sub-figures where each sub-figure contains a bar chart.}
    \label{fig:participant}   
\end{figure*}

\begin{figure*}[h]
    \centering
    \includegraphics[width=\textwidth]{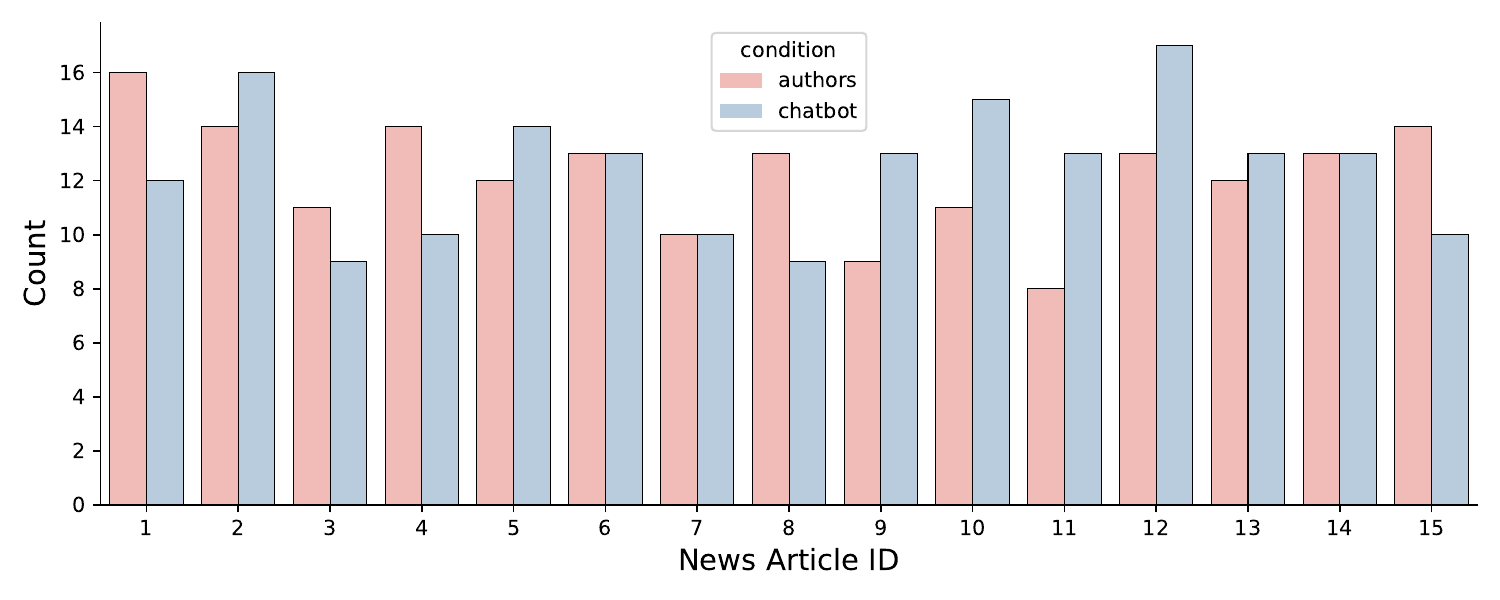}
    \caption{\textbf{Distribution of news articles.} We randomly assigned 15 articles to the participants in the study. Overall, the distribution is uniform in nature.}
    \Description{This figure shows a bar chart with multi-colored bars. }
    \label{fig:article_dist}   
\end{figure*}

\begin{figure*}[h]
    \centering
    \includegraphics[width=0.95\textwidth]{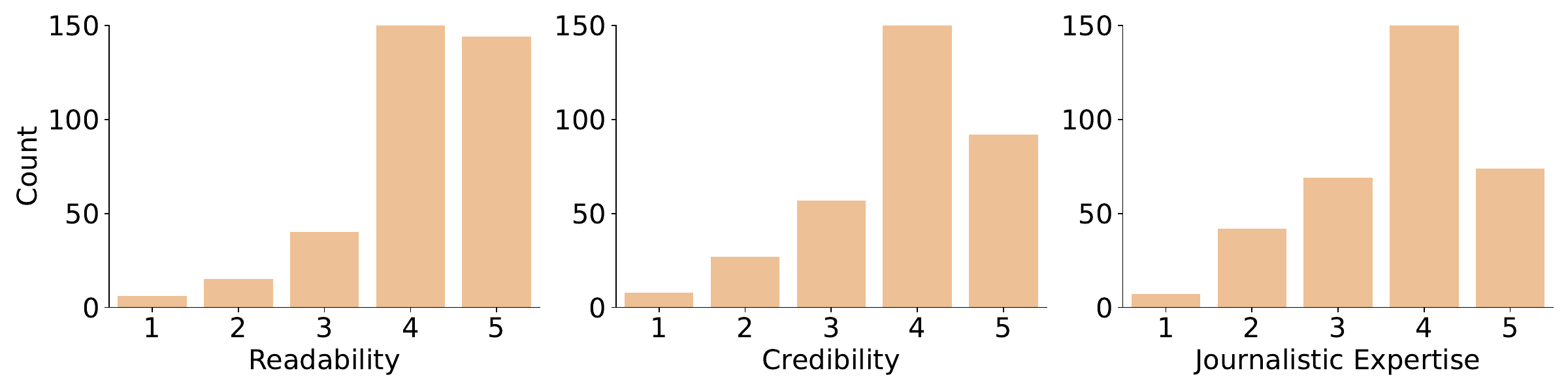}
    \caption{\textbf{Perceived Quality of the news articles.} Participants rated each article on a scale of 1 to 5 across three dimensions: Readability, Credibility, and Journalistic Expertise~\cite{doi:10.1177/1464884916641269}. Y-axis shows counts for each category. Higher values indicate better ratings. }
    \label{fig:quality}
    \Description{This figure has three sub-figures with bar charts with colored bars.}
\end{figure*}

% \section{Appendix B}

\end{document}